\documentclass[
  a4paper,
  aps,
  superscriptaddress,
  twocolumn,
  prx,
]
{revtex4-2}
\usepackage{lineno}
\usepackage{graphicx}  
\usepackage{dcolumn}   
\usepackage{bm}        
\usepackage{amssymb}   
\usepackage{amsmath,stmaryrd}
\usepackage{blkarray, multirow, graphicx, diagbox, color, xcolor, colortbl}
\usepackage{bbm, bbold}
\usepackage{glossaries}
\usepackage{hyphenat}
\usepackage{ifthen}
\usepackage{xkeyval}
\usepackage{moreverb}
\usepackage{rotating}
\usepackage{wrapfig}
\usepackage{slashbox}
\usepackage{xspace}
\usepackage{nicefrac}
\usepackage[]{units}
\usepackage{physics}
\usepackage{booktabs}
\usepackage{braket}
\usepackage[inline]{enumitem}
\usepackage{tabto}
\usepackage{listings}
\usepackage{xstring}
\usepackage{lipsum}
\usepackage{scalerel}

\def\ReplaceStr#1{%
	\IfSubStr{#1}{p}{%
		\StrSubstitute{#1}{p}{.}}{#1}}

\usepackage[caption=false]{subfig}
\newcommand\subfigref[1]{\protect\subref{#1}}

\usepackage[customcolors]{hf-tikz} 
\usepackage{tikz}
\usepackage{calc}
\usetikzlibrary{external}
\usepackage{pgffor}
\usepackage{pgfplots}
\pgfplotsset{compat=1.13}
\usepackage{pgfplotstable}
\usepgfplotslibrary{groupplots}
\usepgfplotslibrary{fillbetween}
\usepgfplotslibrary{colorbrewer}

\tikzstyle{n} = [draw,shape=ellipse,minimum size=1.5em,inner sep=0pt,fill=white!20, minimum width=2.5em]
\tikzstyle{Init} = [n,color=green,fill=green!20,text=black]
\tikzstyle{Fin} = [n,color=red,fill=red!20,text=black]
\tikzstyle{Ghost} = [minimum size=1.5em,inner sep=0pt,color=white,text=black]
\tikzstyle{Multiple} = [draw,shape=rect,minimum size=2em,inner sep=0pt]

\tikzstyle{ghostA} = [text=red!70,thick, minimum size=2*(5pt-\pgflinewidth), inner sep=0pt, outer sep=0pt]
\tikzstyle{ghostB} = [text=blue!70,thick, minimum size=2*(5pt-\pgflinewidth), inner sep=0pt, outer sep=0pt]
\tikzstyle{siteA} = [regular polygon, regular polygon sides=3, shape border rotate= 30, draw=red!50,fill=red!20,thick,inner sep=0pt,minimum width=1.5em,font=\footnotesize]
\tikzstyle{siteB} = [regular polygon, regular polygon sides=3, shape border rotate= -30, draw=green!50,fill=green!20,thick,inner sep=0pt,minimum width=1.5em,font=\footnotesize]
\tikzstyle{op} = [regular polygon, regular polygon sides=4, draw=orange!50, fill=orange!20, thick, inner sep=0.2pt, minimum width=0.25em, minimum height=0.5em,font=\footnotesize]
\tikzstyle{gate} = [rectangle, rounded corners=2pt, draw=orange!50, fill=orange!20, thick, inner sep=0.2pt, minimum width=0.25em, minimum height=0.5em,font=\footnotesize]
\tikzstyle{opghost} = [regular polygon, regular polygon sides=4, thick, inner sep=0.2pt, minimum width=1.25em, minimum height=1.5em,font=\footnotesize]
\tikzstyle{site} = [circle,draw=blue!50,fill=blue!20,thick,inner sep=0.2pt,minimum width=1.25em,font=\footnotesize]
\tikzstyle{hiddensite} = [circle,draw=white!50,fill=white!20,thick,inner sep=0.2pt,minimum width=1.25em,font=\footnotesize]
\tikzstyle{nosite} = [circle,draw=white,fill=white,thick,inner sep=0.1pt,minimum width=1.5em]
\tikzstyle{ghost} = [font=\footnotesize]
\tikzstyle{intersite} = [regular polygon, regular polygon sides=4, shape border rotate= 45, draw=black!50,fill=black!20,thick,inner sep=0pt,minimum width=1.5em]
\tikzstyle{ld} = [inner sep=1pt, font=\small]
\tikzstyle{unsite} = [circle, outer sep=0pt,inner sep=0.2pt,minimum width=1.25em]

\definecolor{colorA}{rgb} {0.58,0,0.8275}
\definecolor{colorB}{rgb} {0.11,0.663,0.51}
\definecolor{colorC}{rgb} {0.3373,0.7059,0.9137}
\definecolor{colorD}{rgb} {0.902,0.6235,0}
\definecolor{colorE}{rgb} {0.9451,0.902,0.3255}
\definecolor{colorF}{rgb} {0.3373,0.3255,0.902}
\definecolor{colorG}{rgb} {0.9451,0.3255,0.3373}
\definecolor{cbColorA}{HTML} {2D4C8D}
\definecolor{cbColorB}{HTML} {EB821D}
\definecolor{cbColorC}{HTML} {C3342F}
\definecolor{cbColorD}{HTML} {24451B}
\definecolor{cbColorE}{HTML} {2C0041}
\definecolor{cbColorF}{HTML} {D9D9D9}

\definecolor{stBlue}{HTML} {5566AA}
\definecolor{stGreen}{HTML} {117733}
\definecolor{stCyan}{HTML} {33BBEE}
\definecolor{stTeal}{HTML} {009988}
\definecolor{stOrange}{HTML} {EE7733}
\definecolor{stYellow}{HTML} {F7F056}
\definecolor{stRed}{HTML} {CC3311}
\definecolor{stMagenta}{HTML} {EE3377}
\definecolor{stGrey}{HTML} {BBBBBB}

\usetikzlibrary
{
	calc,
	decorations,
	plotmarks,
	patterns,
	positioning,
	petri,
	arrows,
	decorations.markings,
	backgrounds,
	fit,
	graphs,
	shapes.geometric,
	decorations.pathmorphing,
	shapes.misc,
	shapes,
	tikzmark,
	pgfplots.colorbrewer,
	fpu,
	svg.path
}

\usepackage{ocgx}

\usetikzlibrary{ocgx}

\pgfplotsset{
        cycle from colormap manual style/.style={
            x=3cm,y=10pt,ytick=\empty,
            stack plots=y,
            every axis plot/.style={line width=2pt},
        },
}

\tikzset{>=stealth}

\tikzset{->-/.style={decoration={
			markings,
			mark=at position .5 with {\arrow{>}}},postaction={decorate}}}

\tikzset{-<-/.style={decoration={
			markings,
			mark=at position .5 with {\arrow{<}}},postaction={decorate}}}

\tikzstyle{orientedsnake} = [
decorate, 
decoration={snake},
->
]  
\tikzstyle{orientedshortarrow} = [
decoration={markings,
	mark=at position .33 with {\arrow{>}}},
postaction={decorate}
]  
\tikzstyle{orientedlongarrow} = [
decoration={markings,
	mark=at position .67 with {\arrow{>}}},
postaction={decorate}
]
\tikzset{dbl/.style={double,
		double equal sign distance,
		-implies,
		shorten >=10pt,
		shorten <=10pt}}
\tikzset{
	between/.style args={#1 and #2}{
		at = ($(#1)!0.5!(#2)$)
	}
}

\tikzstyle{process} = [rectangle, minimum width=3cm, minimum height=1cm, text centered, text width=5cm, draw=black]
\tikzstyle{io} = [trapezium, trapezium left angle=70, trapezium right angle=110, minimum width=3cm, minimum height=1cm, text centered, text width=7cm, draw=black]
\tikzstyle{choose} = [diamond, inner sep=1pt, minimum width=2cm, minimum height=2cm, text centered, text width=1.5cm, draw=black]
\tikzstyle{arrow} =[thick,->, >=stealth]

\pgfmathdeclarefunction{peierlspotential}{6}%
{
	\pgfmathparse
	{
		#2 / (#3 * #5) 
		* sin(deg(#1 * #3 - #3 * #5 * (x)))
		* exp(-( ( #1 - #5 * ((x) - #4) ) )^2.0 / ( #6^2.0 ) )
	}%
}

\pgfmathdeclarefunction{linearFct}{2}%
{%
	\pgfmathparse{#1*x+#2}%
}
\pgfmathdeclarefunction{quadFct}{3}%
{%
	\pgfmathparse{#1*x*x+#2*x+#3}%
}
\pgfmathdeclarefunction{logFct}{2}%
{%
	\pgfmathparse{#1*log10(x)+#2}%
}
\pgfmathdeclarefunction{expFct}{2}%
{%
	\pgfmathparse{#1*exp(-x/#2)}%
}

\newboolean{buildCurrentBlockInline}
\setboolean{buildCurrentBlockInline}{false}
\newboolean{rebuildCurrentBlock}
\setboolean{rebuildCurrentBlock}{false}

\newboolean{rebuildData}
\setboolean{rebuildData}{false}

\newboolean{removeData}
\setboolean{removeData}{false}

\graphicspath{{figures/}}

\newboolean{buildtikzpics}
\setboolean{buildtikzpics}{false}
\newif\ifrebuildtikz
\newif\ifChangeMode
\ChangeModetrue
\ChangeModefalse
\ifthenelse{\boolean{buildtikzpics}}
{
	\rebuildtikztrue
	\usetikzlibrary{external}
	\tikzexternalize[optimize=false,prefix=figures/autogen/]%
}
{
	\rebuildtikzfalse
}

\usepackage[normalem]{ulem}
\usepackage[colorlinks,bookmarks=false,citecolor=blue,linkcolor=black,urlcolor=blue]{hyperref}
\usepackage{cleveref}
\Crefname{appendix}{Appendix}{Appendices}
\Crefname{equation}{Equation}{Equations}
\Crefname{figure}{Figure}{Figures}
\Crefname{section}{Section}{Sections}
\Crefname{tabular}{Tabular}{Tabulars}
\crefname{appendix}{App.}{Apps.}
\crefname{equation}{Eq.}{Eqs.}
\crefname{figure}{Fig.}{Figs.}
\crefname{section}{Sec.}{Secs.}
\crefname{tabular}{Tab.}{Tabs.}

\ExplSyntaxOn
\DeclareExpandableDocumentCommand \eval { m } { \fp_eval:n { #1 } }
\ExplSyntaxOff

\makeatletter
\def\pgfplotsutil@decstringcounter#1{%
 \begingroup
  \c@pgf@counta=#1\relax
  \advance\c@pgf@counta by -1
  \edef#1{\the\c@pgf@counta}%
  \pgfmath@smuggleone#1%
 \endgroup
}%

\pgfplotsset{
/pgfplots/each nth point*/.style 2 args={%
/pgfplots/x filter/.append code={%
 \ifnum\coordindex=0
  \def\c@pgfplots@eachnthpoint@xfilter{0}%
  \edef\c@pgfplots@eachnthpoint@xfilter@cmp{#1}%
 \else
  \ifnum\coordindex>#2\relax
   \pgfplotsutil@advancestringcounter\c@pgfplots@eachnthpoint@xfilter
   \ifx\c@pgfplots@eachnthpoint@xfilter@cmp\c@pgfplots@eachnthpoint@xfilter
    \def\c@pgfplots@eachnthpoint@xfilter{0}%
   \else
    \let\pgfmathresult\pgfutil@empty
   \fi
  \fi
 \fi
}%
},
}
\makeatother

\newcommand{\printpgfnumberorder}[1]%
{%
	\pgfmathfloatparsenumber{#1}%
	\pgfmathfloattomacro{\pgfmathresult}{\Fn}{\Mn}{\En}%
	\pgfmathparse{\Fn==2 ? "-" : ""}%
	\edef\Sn{\pgfmathresult}%
	\Sn 10^{\En}%
}

\newcounter{marknumber}
\pgfplotsset{
	error bars/every nth mark/.style={
		/pgfplots/error bars/draw error bar/.prefix code={
			\pgfmathtruncatemacro\marknumbercheck{mod(floor(\themarknumber/2),#1)}
			\ifnum\marknumbercheck=0
			\else
			\begin{scope}[opacity=0]
				\fi
			},
			/pgfplots/error bars/draw error bar/.append code={
				\ifnum\marknumbercheck=0
				\else
			\end{scope}
			\fi
			\stepcounter{marknumber}    
		}
	}
}

\pgfplotsset{
    /pgf/declare function={
        vk(\x,\a,\b,\c) = \c*sqrt(1.0-(-2*cos(\x)-\b)/sqrt((-2*cos(\x)-\b)^2+\a^2));
    },
}

\pgfplotsset{
/pgfplots/colormap={magma}{
	rgb255=(0,0,4) rgb255=(0,0,6) rgb255=(1,0,7) rgb255=(1,1,9) rgb255=(1,1,11) rgb255=(2,2,13) rgb255=(2,2,15) rgb255=(3,3,17) rgb255=(4,3,19) rgb255=(4,4,21) rgb255=(5,4,23) rgb255=(6,5,25) rgb255=(7,5,27) rgb255=(8,6,29) rgb255=(9,7,31) rgb255=(10,7,34) rgb255=(11,8,36) rgb255=(12,9,38) rgb255=(13,10,40) rgb255=(14,10,42) rgb255=(15,11,44) rgb255=(16,12,47) rgb255=(17,12,49) rgb255=(18,13,51) rgb255=(20,13,53) rgb255=(21,14,56) rgb255=(22,14,58) rgb255=(23,15,60) rgb255=(24,15,63) rgb255=(26,16,65) rgb255=(27,16,68) rgb255=(28,16,70) rgb255=(30,16,73) rgb255=(31,17,75) rgb255=(32,17,77) rgb255=(34,17,80) rgb255=(35,17,82) rgb255=(37,17,85) rgb255=(38,17,87) rgb255=(40,17,89) rgb255=(42,17,92) rgb255=(43,17,94) rgb255=(45,16,96) rgb255=(47,16,98) rgb255=(48,16,101) rgb255=(50,16,103) rgb255=(52,16,104) rgb255=(53,15,106) rgb255=(55,15,108) rgb255=(57,15,110) rgb255=(59,15,111) rgb255=(60,15,113) rgb255=(62,15,114) rgb255=(64,15,115) rgb255=(66,15,116) rgb255=(67,15,117) rgb255=(69,15,118) rgb255=(71,15,119) rgb255=(72,16,120) rgb255=(74,16,121) rgb255=(75,16,121) rgb255=(77,17,122) rgb255=(79,17,123) rgb255=(80,18,123) rgb255=(82,18,124) rgb255=(83,19,124) rgb255=(85,19,125) rgb255=(87,20,125) rgb255=(88,21,126) rgb255=(90,21,126) rgb255=(91,22,126) rgb255=(93,23,126) rgb255=(94,23,127) rgb255=(96,24,127) rgb255=(97,24,127) rgb255=(99,25,127) rgb255=(101,26,128) rgb255=(102,26,128) rgb255=(104,27,128) rgb255=(105,28,128) rgb255=(107,28,128) rgb255=(108,29,128) rgb255=(110,30,129) rgb255=(111,30,129) rgb255=(113,31,129) rgb255=(115,31,129) rgb255=(116,32,129) rgb255=(118,33,129) rgb255=(119,33,129) rgb255=(121,34,129) rgb255=(122,34,129) rgb255=(124,35,129) rgb255=(126,36,129) rgb255=(127,36,129) rgb255=(129,37,129) rgb255=(130,37,129) rgb255=(132,38,129) rgb255=(133,38,129) rgb255=(135,39,129) rgb255=(137,40,129) rgb255=(138,40,129) rgb255=(140,41,128) rgb255=(141,41,128) rgb255=(143,42,128) rgb255=(145,42,128) rgb255=(146,43,128) rgb255=(148,43,128) rgb255=(149,44,128) rgb255=(151,44,127) rgb255=(153,45,127) rgb255=(154,45,127) rgb255=(156,46,127) rgb255=(158,46,126) rgb255=(159,47,126) rgb255=(161,47,126) rgb255=(163,48,126) rgb255=(164,48,125) rgb255=(166,49,125) rgb255=(167,49,125) rgb255=(169,50,124) rgb255=(171,51,124) rgb255=(172,51,123) rgb255=(174,52,123) rgb255=(176,52,123) rgb255=(177,53,122) rgb255=(179,53,122) rgb255=(181,54,121) rgb255=(182,54,121) rgb255=(184,55,120) rgb255=(185,55,120) rgb255=(187,56,119) rgb255=(189,57,119) rgb255=(190,57,118) rgb255=(192,58,117) rgb255=(194,58,117) rgb255=(195,59,116) rgb255=(197,60,116) rgb255=(198,60,115) rgb255=(200,61,114) rgb255=(202,62,114) rgb255=(203,62,113) rgb255=(205,63,112) rgb255=(206,64,112) rgb255=(208,65,111) rgb255=(209,66,110) rgb255=(211,66,109) rgb255=(212,67,109) rgb255=(214,68,108) rgb255=(215,69,107) rgb255=(217,70,106) rgb255=(218,71,105) rgb255=(220,72,105) rgb255=(221,73,104) rgb255=(222,74,103) rgb255=(224,75,102) rgb255=(225,76,102) rgb255=(226,77,101) rgb255=(228,78,100) rgb255=(229,80,99) rgb255=(230,81,98) rgb255=(231,82,98) rgb255=(232,84,97) rgb255=(234,85,96) rgb255=(235,86,96) rgb255=(236,88,95) rgb255=(237,89,95) rgb255=(238,91,94) rgb255=(238,93,93) rgb255=(239,94,93) rgb255=(240,96,93) rgb255=(241,97,92) rgb255=(242,99,92) rgb255=(243,101,92) rgb255=(243,103,91) rgb255=(244,104,91) rgb255=(245,106,91) rgb255=(245,108,91) rgb255=(246,110,91) rgb255=(246,112,91) rgb255=(247,113,91) rgb255=(247,115,92) rgb255=(248,117,92) rgb255=(248,119,92) rgb255=(249,121,92) rgb255=(249,123,93) rgb255=(249,125,93) rgb255=(250,127,94) rgb255=(250,128,94) rgb255=(250,130,95) rgb255=(251,132,96) rgb255=(251,134,96) rgb255=(251,136,97) rgb255=(251,138,98) rgb255=(252,140,99) rgb255=(252,142,99) rgb255=(252,144,100) rgb255=(252,146,101) rgb255=(252,147,102) rgb255=(253,149,103) rgb255=(253,151,104) rgb255=(253,153,105) rgb255=(253,155,106) rgb255=(253,157,107) rgb255=(253,159,108) rgb255=(253,161,110) rgb255=(253,162,111) rgb255=(253,164,112) rgb255=(254,166,113) rgb255=(254,168,115) rgb255=(254,170,116) rgb255=(254,172,117) rgb255=(254,174,118) rgb255=(254,175,120) rgb255=(254,177,121) rgb255=(254,179,123) rgb255=(254,181,124) rgb255=(254,183,125) rgb255=(254,185,127) rgb255=(254,187,128) rgb255=(254,188,130) rgb255=(254,190,131) rgb255=(254,192,133) rgb255=(254,194,134) rgb255=(254,196,136) rgb255=(254,198,137) rgb255=(254,199,139) rgb255=(254,201,141) rgb255=(254,203,142) rgb255=(253,205,144) rgb255=(253,207,146) rgb255=(253,209,147) rgb255=(253,210,149) rgb255=(253,212,151) rgb255=(253,214,152) rgb255=(253,216,154) rgb255=(253,218,156) rgb255=(253,220,157) rgb255=(253,221,159) rgb255=(253,223,161) rgb255=(253,225,163) rgb255=(252,227,165) rgb255=(252,229,166) rgb255=(252,230,168) rgb255=(252,232,170) rgb255=(252,234,172) rgb255=(252,236,174) rgb255=(252,238,176) rgb255=(252,240,177) rgb255=(252,241,179) rgb255=(252,243,181) rgb255=(252,245,183) rgb255=(251,247,185) rgb255=(251,249,187) rgb255=(251,250,189) rgb255=(251,252,191)}
}
\newacronym[shortplural={MPS}]{MPS}{MPS}{matrix\hyp product state}
\newacronym{MPSs}{MPS}{matrix\hyp product states}
\newacronym[shortplural={PP-MPS}]{PP-MPS}{PP-MPS}{projected purified matrix\hyp product state}
\newacronym{PP-MPSs}{PP-MPS}{projected purified matrix\hyp product states}
\newacronym{MPO}{MPO}{matrix-product operator}
\newacronym{SVD}{SVD}{singular-value decomposition}
\newacronym{QCS}{QCS}{quantum-computer simulator}
\newacronym{FSM}{FSM}{finite-state machine}
\newacronym{ACA}{ACA}{adaptive cross approximation}
\newacronym{1D}{1D}{one\hyp dimensional}
\newacronym{QC}{QC}{quantum computer}
\newacronym{CDW}{CDW}{charge\hyp density wave}
\newacronym{BOW}{BOW}{bond\hyp order wave}
\newacronym{SDW}{SDW}{spin\hyp density wave}
\newacronym{ARPES}{ARPES}{angle-resolved photoemission spectroscopy}
\newacronym{OBC}{OBC}{open-boundary conditions}
\newacronym{PBC}{PBC}{periodic-boundary conditions}
\newacronym{TEBD}{TEBD}{time-evolving block-decimation}
\newacronym{TDVP}{TDVP}{time\hyp dependent variational principle}
\newacronym{1TDVP}{1TDVP}{single\hyp site time\hyp dependent variational principle}
\newacronym{2TDVP}{2TDVP}{two\hyp site time\hyp dependent variational principle}
\newacronym{iff}{iff}{if and only if}
\newacronym{DFT}{DFT}{density\hyp functional theory}
\newacronym{DMFT}{DMFT}{dynamical mean\hyp field theory}
\newacronym{DMRG}{DMRG}{density\hyp matrix renormalization group}
\newacronym{1DMRG}{1DMRG}{single-site density\hyp matrix renormalization group}
\newacronym{2DMRG}{2DMRG}{two-site density\hyp matrix renormalization group}
\newacronym{DMRG3S}{DMRG3S}{strictly single-site density\hyp matrix renormalization group}
\newacronym{iDMRG}{iDMRG}{inifinite\hyp size density\hyp matrix renormalization group}
\newacronym{tDMRG}{tDMRG}{time\hyp dependent density\hyp matrix renormalization group}
\newacronym{PP-DMRG}{PP-DMRG}{projected purified density\hyp matrix renormalization group}
\newacronym{QMC}{QMC}{quantum Monte Carlo}
\newacronym{AIM}{AIM}{Anderson impurity model}
\newacronym{SIAM}{SIAM}{single impurity Anderson model}
\newacronym{LDA}{LDA}{local\hyp density approximation}
\newacronym{LBNL}{LBNL}{Lawrence Berkeley National Laboratory}
\newacronym{ED}{ED}{exact diagonalization}
\newacronym{QPT}{QPT}{quantum phase transition}
\newacronym{QCP}{QCP}{quantum critical point}
\newacronym{ETH}{ETH}{eigenstate thermalization hypothesis}
\newacronym{EHM}{EHM}{extended Hubbard model}
\newacronym{AKLT}{AKLT}{Affleck\hyp Lieb\hyp Kennedy\hyp Tasaki}
\newglossaryentry{QR}{name={QR},description={QR decomposition}}
\newacronym{TNS}{TNS}{tensor\hyp network state}
\newacronym{SM}{SM}{supplemental material}
\newacronym{NOO}{NOO}{natural orbital occupation}
\newacronym{NO}{NO}{natural orbital}
\newacronym{LRO}{LRO}{long\hyp range order}
\newacronym{qLRO}{qLRO}{quasi\hyp long\hyp range order}
\newacronym{ODLRO}{ODLRO}{off\hyp diagonal long\hyp range order}
\newacronym{SC}{SC}{Superconductivity}
\newacronym{VBGS}{VBGS}{valence bond ground-state}
\newacronym{PEPS}{PEPS}{projected entangled pair\hyp states}
\newacronym{ALS}{ALS}{alternating least squares}
\newacronym{BdG}{BdG}{Bogoljubov de-Gennes}
\newacronym{TFIM}{TFI}{transverse field Ising model}
\newacronym{PP}{PP}{projected purification}
\newacronym{BEC}{BEC}{Bose\hyp Einstein condensate}
\newacronym{JWT}{JWT}{Jordan\hyp Wigner transformation}
\newacronym{NISQ}{NISQ}{noisy intermediate scale quantum}
\newacronym{NN}{NN}{nearest\hyp neighbor}
\newacronym{NNN}{NNN}{next\hyp nearest\hyp neighbor}
\newacronym{SPDM}{SPDM}{single\hyp particle density matrix} 
\newacronym{HCB}{HCB}{hardcore bosons}
\newacronym{SF}{SF}{spinless fermions}
\newacronym{fRG}{fRG}{functional renormalization group}
\newacronym{LE}{LE}{Luther\hyp Emery}
\newacronym{ASP}{ASP}{adiabatic state preparation}
\newacronym{VQE}{VQE}{variational quantum eigensolver}
\newacronym{METTS}{METTS}{minimally\hyp entangled typical thermal states}
\newacronym{SSH}{SSH}{Su\hyp Schrieffer\hyp Heeger}
\newacronym{GSE}{GSE}{global subspace\hyp expansion}
\newacronym{LSE}{LSE}{local subspace\hyp expansion}
\newacronym{2RDM}{2RDM}{two\hyp particle reduced density\hyp matrix}
\newacronym{BCS}{BCS}{Bardeen\hyp Cooper\hyp Schrieffer}
\newcommand{\acsaddress}{Department of Physics and Arnold Sommerfeld Center for Theoretical Physics (ASC), Ludwig-Maximilians-Universit\"{a}t M\"{u}nchen, D-80333 Munich, Germany}
\newcommand{\mcqstaddress}{Munich Center for Quantum Science and Technology (MCQST), D-80799 M\"{u}nchen, Germany}

\newcommand{\stanfordaddress}{Department of Physics, Stanford University, Stanford, CA 93405, USA}
\newcommand{\ucsdaddress}{Department of Chemistry and Biochemistry, University of California, San Diego, La Jolla, California 92093, USA}

\newcommand{\nodagger}[0]{{\vphantom{\dagger}}}
\newcommand{\noprime}[0]{{\vphantom{\prime}}}

\newcommand{\overbar}[1]{\mkern 1.5mu\overline{\mkern-1.5mu#1\mkern-1.5mu}\mkern 1.5mu}

\newcommand{\sgn}[0]{\operatorname{sgn}}

\definecolor{lmugreen}{RGB}{0.0, 148, 64}
\definecolor{Gray}{gray}{0.9}
\newif\ifshowcomments\showcommentstrue

\Crefname{appendix}{Appendix}{Appendices}
\Crefname{equation}{Equation}{Equations}
\Crefname{figure}{Figure}{Figures}
\Crefname{section}{Section}{Sections}
\Crefname{tabular}{Tabular}{Tabulars}
\crefname{appendix}{App.}{Apps.}
\crefname{equation}{Eq.}{Eqs.}
\crefname{figure}{Fig.}{Figs.}
\crefname{section}{Sec.}{Secs.}
\crefname{tabular}{Tab.}{Tabs.}

\definecolor{midnight_blue}{HTML} {002952}
\definecolor{baby_blue} {HTML} {ddeeff}
\definecolor{hot_pink}{HTML} {D15D84}
\definecolor{corn_flower}{HTML} {78A1E5}
\definecolor{mauve}{HTML} {905171}
\definecolor{purple}{HTML} {4C3B55}
\definecolor{tiffany_blue}{HTML} {6AA4B0}
\definecolor{gunmetal_grey}{HTML} {4C5355}
\definecolor{honeysuckle}{HTML} {C44B4F}
\definecolor{blueish}{HTML} {226E9C}
\definecolor{bluegray}{HTML} {607D86}
\definecolor{cinnabar}{HTML} {E66354}
\definecolor{walnut}{HTML} {4D181C}
\definecolor{mahagony}{HTML} {4B1816}
\definecolor{aegean_blue}{HTML} {144058}
\definecolor{honey}{HTML} {E58D2E}
\definecolor{persimmon}{HTML} {DD671E}
\definecolor{mimosa}{HTML} {D7A449}
\definecolor{lilac}{HTML} {D5CAE4}

\definecolor{darkteal}{HTML} {3A6D80}
\definecolor{amber}{HTML} {F3CD53}
\definecolor{squash}{HTML} {D56729}
\definecolor{vermillion}{HTML} {9D402D}
\definecolor{cascades_green}{HTML} {355952}
\definecolor{baby_pink}{HTML} {FDC2E4}
\definecolor{wisteria}{HTML} {D3C5E5}
\definecolor{burgundy}{HTML} {800020}

\definecolor{shinyblue}{HTML} {257eee}
\definecolor{fatyellow}{HTML} {fbfe36}
\definecolor{glowyred}{HTML} {ea1822}

\colorlet{cbBegin}{baby_blue}
\colorlet{cbEnd}{midnight_blue}

\colorlet{colorUOpO}{midnight_blue}
\colorlet{colorUOpXXV}{persimmon}
\colorlet{colorUOpL}{burgundy}
\colorlet{colorUOpLXXV}{cascades_green}

\definecolor{orcidlogocol}{HTML}{A6CE39}
\tikzset{
  orcidlogo/.pic={
    \fill[orcidlogocol] svg{M256,128c0,70.7-57.3,128-128,128C57.3,256,0,198.7,0,128C0,57.3,57.3,0,128,0C198.7,0,256,57.3,256,128z};
    \fill[white] svg{M86.3,186.2H70.9V79.1h15.4v48.4V186.2z}
                 svg{M108.9,79.1h41.6c39.6,0,57,28.3,57,53.6c0,27.5-21.5,53.6-56.8,53.6h-41.8V79.1z M124.3,172.4h24.5c34.9,0,42.9-26.5,42.9-39.7c0-21.5-13.7-39.7-43.7-39.7h-23.7V172.4z}
                 svg{M88.7,56.8c0,5.5-4.5,10.1-10.1,10.1c-5.6,0-10.1-4.6-10.1-10.1c0-5.6,4.5-10.1,10.1-10.1C84.2,46.7,88.7,51.3,88.7,56.8z};
  }
}

\newcommand\orcidicon[1]{\href{https://orcid.org/#1}{\mbox{\scalerel*{
  \tikzset{external/export=true}
  \begin{tikzpicture}[yscale=-1, transform shape]
    \pic{orcidlogo};
  \end{tikzpicture}
}{|}}}}
\begin{document}
\title{Cooper\hyp Paired Bipolaronic Superconductors}
\author{M.~Grundner\hspace{0.069cm}\orcidicon{0000-0002-7694-0053}}
\affiliation{\acsaddress}
\affiliation{\mcqstaddress}
\author{T.~Blatz\hspace{0.069cm}\orcidicon{0009-0000-3988-2176}}
\affiliation{\acsaddress}
\affiliation{\mcqstaddress}
\author{J.~Sous\hspace{0.069cm}\orcidicon{0000-0002-9994-5789}}
\affiliation{\stanfordaddress}
\affiliation{\ucsdaddress}
\author{U.~Schollw\"ock\hspace{0.069cm}\orcidicon{0000-0002-2538-1802}}
\affiliation{\acsaddress}
\affiliation{\mcqstaddress}
\author{S.~Paeckel\hspace{0.069cm}\orcidicon{0000-0001-8107-069X}}
\affiliation{\acsaddress}
\affiliation{\mcqstaddress}
\date{\today}
\begin{abstract}
Light\hyp mass bipolarons in off\hyp diagonally coupled electron\hyp phonon systems provide a potential route to bipolaronic high\hyp $T_\mathrm{c}$ superconductivity.
While there has been numerical progress in the physically relevant limit of slow phonons, more insights are needed to fully understand to what extent this mechanism survives at finite densities and in different regimes.
We address these questions using advanced tensor\hyp network methods applied to the~\acrlong{SSH} model.
Studying both the spectral properties of isolated bipolarons as well as the pairing correlations at small finite densities, we find evidence that the conventional picture of bipolarons as molecular bound states which undergo Bose\hyp Einstein condensation may need to be reconsidered.
Instead, our findings suggest correlation\hyp driven formation of a fragmented condensate with spatially separated polaron pairs, stabilized by strong repulsive electron\hyp electron interactions at moderate values of the electron\hyp phonon coupling.
These spatially modulated charge clouds exhibit pairing with a typical length\hyp scale set by the Fermi momentum.
\end{abstract}
\maketitle
\paragraph*{Introduction\label{sec:introduction}}
Phonon\hyp mediated high\hyp $T_\mathrm{c}$ superconductivity at ambient pressure is a fascinating long\hyp standing theoretical idea challenged by the fact that local (Holstein or Fr\"ohlich) coupling to phonons results in an exponential increase in the quasiparticle mass, suppressing $T_c$~\cite{Froehlich_1950,Froehlich_1954,Holstein_I,Holstein_II,Alexandrov1981,Alexandrov1981b,Alexandrov1986,Chakraverty1998}.
Nevertheless, recent studies on systems with off\hyp diagonal electron\hyp phonon interaction reported the formation of light\hyp mass, small\hyp size bipolarons~\cite{Sous2018,Carbone2021,Sous_2022,ly2023,kim2023,Zhang2023,cai2023} and estimated large values of $T_c$.
This analysis is based on the assumption that a gas of bipolarons would undergo a transition to a~\gls{BEC} at finite densities and that this is not preempted by a competing phase.
Numerical simulations of these models are challenging even at $T=0$, for the physically relevant regime of small phonon frequencies $\Omega$ with adiabaticity ratios $A = \Omega/2t \ll 1$, where $t$ denotes the electronic hopping amplitude.
Here, we use projected purified~\gls{MPS}\hyp based techniques~\cite{schollwoeck_2011,Paeckel2019,Yang_2020prb,Yang_2020,Koehler2021,Stolpp2021} to examine the validity of the assumptions at the example of the prototypical one\hyp dimensional~\gls{SSH} model in the adiabatic limit.
%
%
Our results indicate that in contrast to the anti\hyp adiabatic limit $A > 1$~\cite{Sous2018}, in one dimension isolated bipolarons are unstable in the presence of finite electron\hyp electron interactions.
As a consequence, at finite densities we find that the picture of Bose\hyp Einstein condensation of weakly\hyp interacting molecular bound states does not hold.
In turn, we observe the formation of states with dominant Cooper-pairing.
Due to strong correlations, these states do not condense into a macroscopically occupied state~\cite{Mueller2006}.
Instead, they form a maximally fragmented condensate, a scenario which has been investigated in striped superconductors, only recently, and which suggests a coupling between charge\hyp density modulations with different wave vectors and the superconducting order parameter field~\cite{Wietek2022,Baldelli2023}.
Our findings indicate that dilute superconductors exihibt non\hyp trivial, strongly\hyp correlated ground states formed of spatially coherent polaron pairs~\cite{Alexandrov1981,Alexandrov1981b}.
\paragraph*{Model and Methods\label{sec:methods}}
In the adiabatic limit, the energy cost $\Omega$ to excite phonons is much smaller than the electronic bandwidth, limiting analytical progress.
Consequently, unbiased numerical tools are needed to simulate the spectral properties of polarons and bipolarons.
There have been significant advances based on various numerical techniques~\cite{Weber2015,Weber2016,Weber2017,Shi2018,Weber2020,Jansen2020,Wang2020,Han2020,Weber2022,Jansen2022,Carbonne2022,Goetz2022,li2022suppressed}.
While most of them have been applied successfully to local Holstein or Fr\"ohlich type electron\hyp phonon interactions, here we study the~\gls{SSH}\hyp Hubbard model in one dimension, given by the Hamiltonian
\begin{align}
  \hat H
  &=
  -\sum_{j,\sigma} \left(\hat c^\dagger_{j,\sigma}\hat c^\nodagger_{j+1,\sigma} + \mathrm{h.c.} \right) \left(t - \alpha\hat V^\nodagger_{j, j+1} \right) \notag \\
  &\phantom{=}\quad
  +U\sum_j \hat n^\nodagger_{j,\uparrow} \hat n^\nodagger_{j,\downarrow} + \Omega\sum_j \hat a^\dagger_j \hat a^\nodagger_j \; , \label{eq:ssh-model}
\end{align}
where $\hat c^{(\dagger)}_{j,\sigma}$ are the spin\hyp $\sigma$ electronic ladder operators and $\hat a^{(\dagger)}_j$ denote the annihilation (creation) operators of optical phonons with frequency $\Omega$.
The interaction is given by $\hat V_{j, j+1} = \sqrt{2 M \Omega} \left( \hat X^\nodagger_j - \hat X^\nodagger_{j+1} \right) = \hat a^\dagger_j + a^\nodagger_j - \hat a^\dagger_{j+1} - a^\nodagger_{j+1}$ with the phonon displacements $\hat X^\nodagger_j$, and we choose the oscillator mass $M = 1$.
A relevant energy scale is set by the dimensionless, effective electron\hyp phonon coupling $\lambda=\frac{2\alpha^2}{\Omega t}$.
But care must be taken when tuning $\lambda$.
Since~\cref{eq:ssh-model} is obtained by neglecting higher order couplings, one must restrict the model to the linear approximation in which the displacements do not induce a sign change of the hopping integral~\cite{Barford_2006,Nocera_2021}; we have confirmed that this is the case in our simulations~\cite{supp_mat}.
We use \gls{MPS}~\cite{white_1992,white_1993,schollwoeck_2005,schollwoeck_2011,Paeckel2019} representations tailored to describe large local Hilbert spaces~\cite{Koehler2021,Stolpp2021}.
In order to calculate spectral functions, we combine those with improved time\hyp evolution schemes~\cite{Yang_2020,Yang_2020prb,supp_mat}.
We obtain the isolated (bi\hyp)polaron spectral functions~\cite{Mello1997,Zhang1999} from the time\hyp dependent Greens functions
\begin{align}
  G_\mathrm{P}(j,t) &= \braket{\varnothing | \hat c^\nodagger_{j,\sigma}(t) \hat c^\dagger_{\nicefrac{L}{2},\sigma}(0) | \varnothing}\quad\text{polaron,} \label{eq:gf:polaron}\\
  G_\mathrm{BP}(j,t) &= \braket{\varnothing | \hat d^\nodagger_j(t) \hat d^\dagger_{\nicefrac{L}{2}}(0) | \varnothing}\quad\text{bipolaron,} \label{eq:gf:bipolaron}
\end{align}
where $\ket{\varnothing}$ denotes the combined electron\hyp phonon vacuum state and $\hat d^{(\dagger)}_j = (\hat c^\nodagger_{j,\downarrow}\hat c^\nodagger_{j,\uparrow})^{(\dagger)}$ the two\hyp particle electronic operators.
Taking two Fourier transformations yields the momentum resolved spectral functions $S_\alpha(k,\omega)$ with $\alpha=\mathrm{P,BP}$.
We complement the investigations of isolated bipolarons by studying the ground state of the~\gls{SSH} model at small electronic fillings $n=N/2L \sim 0.1$.
Here, the opening of a spin gap turns the ground state into a Luther\hyp Emery liquid~\cite{Mathey2007,Xianlong2007} with dominant pairing correlations.
An in\hyp depth analysis of the phase diagram at small densities is beyond the scope of this work; here we focus on a careful characterization of the electron pairing.
%
%
For that purpose, we investigate the~\gls{2RDM} in the spin\hyp singlet pairing\hyp channel
\begin{equation}
  \hat\rho^{(2)} = \sum_{ijmn} \rho_{(ij), (mn)} \hat c^\dagger_{i,\uparrow}\hat c^\dagger_{j,\downarrow}\hat c^\nodagger_{m,\downarrow}\hat c^\nodagger_{n,\uparrow} \label{eq:2rdm} \;,
\end{equation}
with the \gls{2RDM} matrix elements $\rho_{(ij),(mn)} = \braket{\psi|\hat c^\dagger_{i,\uparrow}\hat c^\dagger_{j,\downarrow}\hat c^\nodagger_{m,\downarrow}\hat c^\nodagger_{n,\uparrow}|\psi}$.
Note that for the case of Bose\hyp Einstein condensation, one typically studies only the on\hyp site contributions $\rho^\mathrm{BEC}_{(ij),(mn)} = \rho_{(ij),(mn)} \delta_{ij} \delta_{mn}$.
The full algorithmic details as well as investigations of model instabilities are provided in~\cite{supp_mat}.
\paragraph*{Spectral properties of (bi\hyp)polarons in the \gls{SSH} model\label{sec:results}}
\begin{figure}[h]
  \subfloat[\label{fig:spec:bipolaron:u-0.0}]{
    \centering
    \hspace{0.05\textwidth}
    \hspace{-6mm}
    \includegraphics[width=0.45\textwidth]{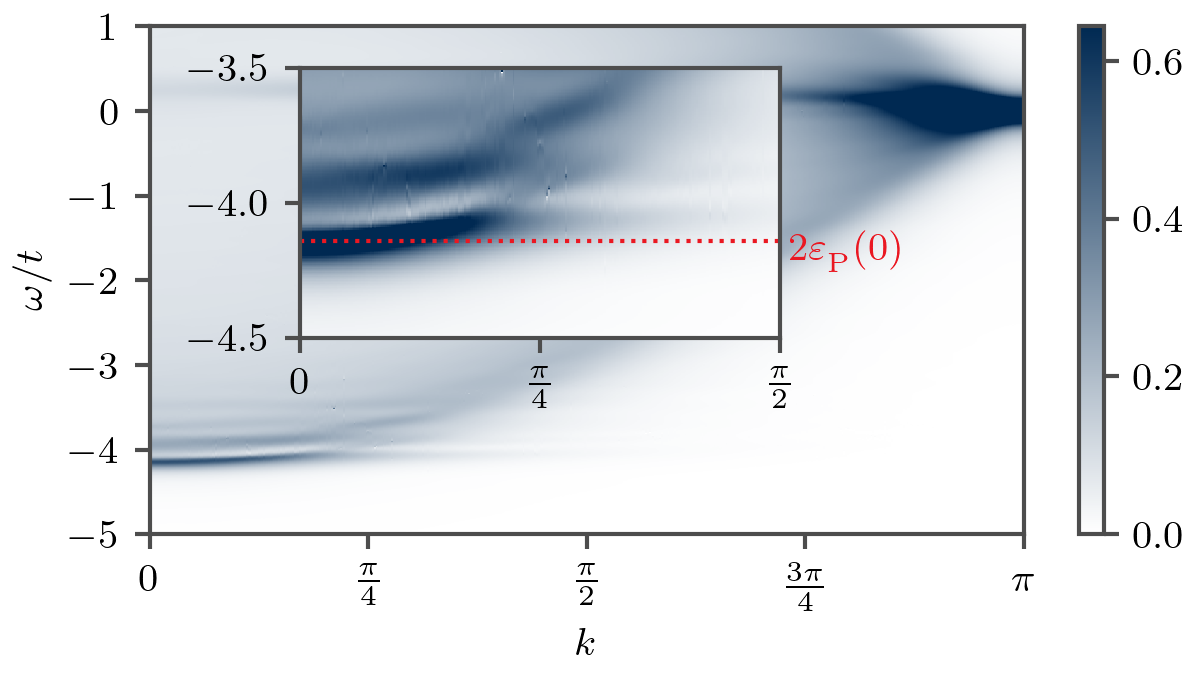}
  }\vspace*{-0.75em}%
  \\
  \subfloat[\label{fig:spec:bipolaron:u-0.5}]{
    \centering
    \hspace{0.05\textwidth}
    \hspace{-6mm}
    \includegraphics[width=0.45\textwidth]{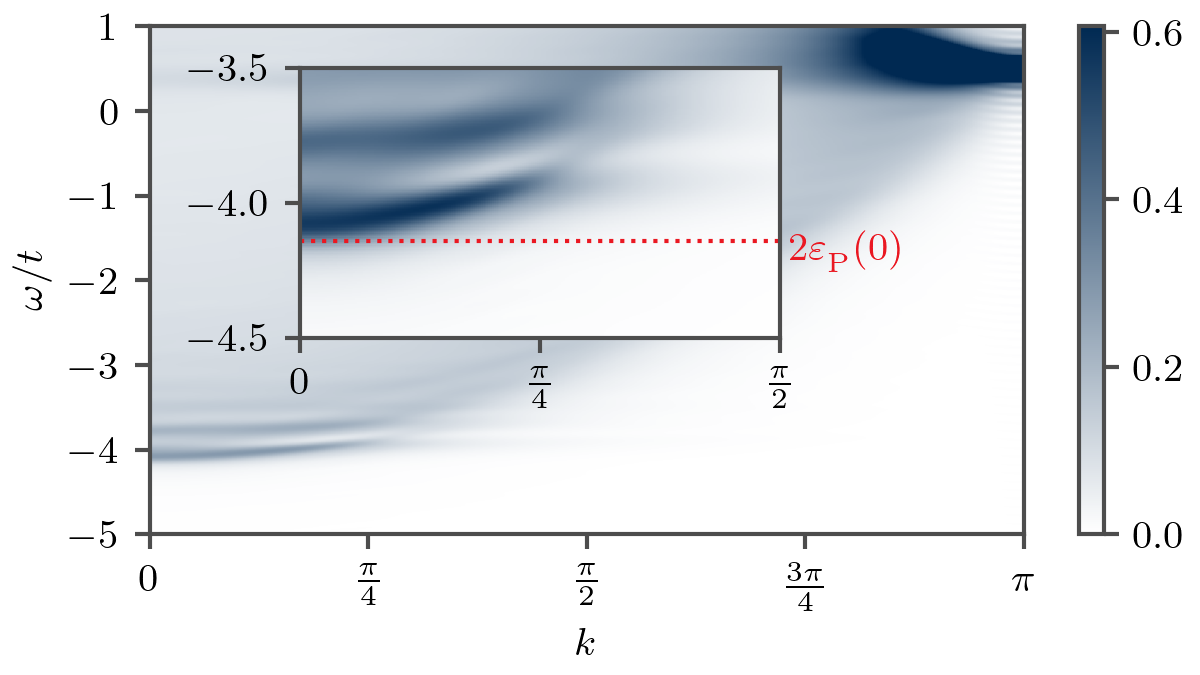}
  }\vspace*{-0.25em}%
  \\
  \subfloat[\label{fig:binding-energies:bipolarons:w-0.2}]{
    \centering
    \includegraphics{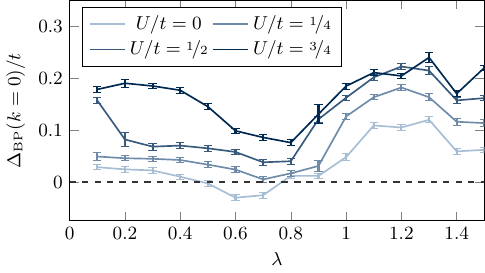}
  }
  \caption{
    \label{fig:spec:bipolaron}
    In \subfigref{fig:spec:bipolaron:u-0.0} we show the pair spectral function for $U/t=0, \Omega/t=0.2 \text{ and } \lambda=0.5$, employing a Lorentzian broadening $\eta/t = 0.05$.
    In the inset we can identify the lowest bipolaronic bound state.
    We also marked the two\hyp polaron ground\hyp state energy $2\varepsilon^\noprime_\mathrm{P}(k=0)$ as dotted line.
    In \subfigref{fig:spec:bipolaron:u-0.5} we display the case $U/t=0.5$.
    Here, the bipolaronic spectral function exhibits an incoherent band, i.e., the bipolarons are unstable towards decaying into two\hyp polaron states.
    In \subfigref{fig:binding-energies:bipolarons:w-0.2}, the extracted bipolaronic binding energies $\Delta_\mathrm{BP}(k) = 2\varepsilon^\noprime_\mathrm{BP}(k) - \varepsilon^\noprime_\mathrm{P}(k)$ as a function of electron\hyp phonon coupling $\lambda$ are shown.
    We observe stable bipolarons only for $U/t=0$ in a small range of electron\hyp phonon couplings $\lambda \in [0.5,0.7]$.
  }
\end{figure}
We begin our analysis by studying the quasi\hyp particle dispersion relations of isolated polarons and bipolarons, extracted from the spectral functions $S_\alpha(k,\omega)$.
We show two\hyp electron spectral functions in~\cref{fig:spec:bipolaron}, calculated for a system of $L=256$ lattice sites.
Fitting the position of maxima in line\hyp cuts at constant momentum $k$ and in the lowest peak of $S_\mathrm{P/BP}(k,\omega)$~\cite{supp_mat}, we determine the quasi\hyp particle dispersion relation $\varepsilon_\mathrm{P/BP}(k)=\varepsilon_\mathrm{P/BP}^0+\frac{k^2}{2m^\noprime_\mathrm{eff}}$ near $k=0$ with the quasi\hyp particle effective mass $m^\noprime_\mathrm{eff} = \left. \frac{\partial^2 E}{\partial k^2} \right\vert_{k=0}$ and the ground\hyp state energy $\varepsilon_\mathrm{P/BP}^0$.
Introducing the bipolaronic binding energy
\begin{equation}
  \Delta_\mathrm{BP}(k) = \varepsilon^\noprime_\mathrm{BP}(k) - 2 \varepsilon^\noprime_\mathrm{P}(k) \;, \label{eq:binding-energy}
\end{equation}
we call two\hyp electron states stable bipolarons if $\Delta_\mathrm{BP}(k)<0$.
Note that for all values of $\lambda$ investigated, we find the minimum of the polaron dispersion relation to be located at $k_\mathrm{P} = 0$, i.e., deep in the adiabatic limit we do not observe a transition to a finite\hyp momentum polaron ground state for $\lambda \leq 1.5$~\cite{Berciu2006,Goodvin2008,Berciu2010,Marchand2010}.
First, we note in~\cref{fig:spec:bipolaron:u-0.0,fig:spec:bipolaron:u-0.5} a strong feature at $k=\pi$, which can be attributed to unstable, localized on\hyp site bipolaron\hyp like states~\cite{supp_mat}.
The extracted bipolaronic binding energies are shown in~\cref{fig:binding-energies:bipolarons:w-0.2} varying both the electron\hyp phonon and the Hubbard interaction.
Strikingly, we do not observe stable bipolarons except for a small region $\lambda \in [0.5,0.7]$ for non\hyp interacting electrons $U/t=0$.
\Cref{fig:spec:bipolaron:u-0.5} illustrates this finding for the case of $U/t=0.5$: There is only a broad, incoherent band around $\omega/t\sim -4.1$ indicating the decay of bipolarons into polarons that form a continuum, marked by the dashed, red line in the inset.
%
%
%
Nevertheless, the overall energy differences between polaronic and bipolaronic ground states are very small $\Delta_\mathrm{BP}/t \sim 0.1$, indicating that at finite fillings correlation effects may play a crucial role.
Note that in order to connect to previous works we also considered the anti\hyp adiabatic regime $\Omega/t=1$.
In this limit, we observe tightly bound bipolarons (shown in~\cite{supp_mat}), in agreement with the findings discussed in~\cite{Sous2018}.
\paragraph*{Maximally fragmented condensation at finite densities}
\begin{figure}
  \centering
  \subfloat[\label{fig:2rdm:eig-vals}]{
    \centering
    \hspace*{-1em}
    \includegraphics{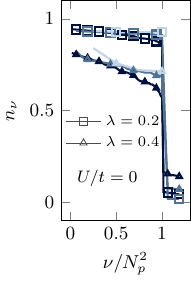}
    \hspace*{-2em}
    \includegraphics{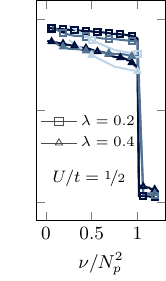}
    \hspace*{-2em}
    \includegraphics{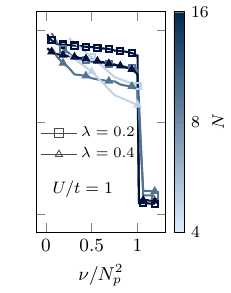}
  }%
  \\
  \subfloat[\label{fig:2rdm:commutators}]{
    \centering
    \hspace*{-1em}
    \includegraphics{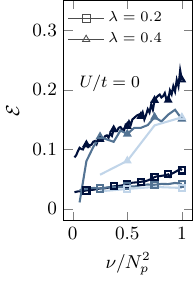}
    \hspace*{-2em}
    \includegraphics{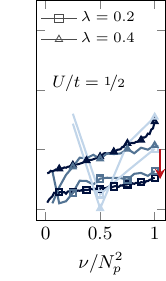}
    \hspace*{-2em}
    \includegraphics{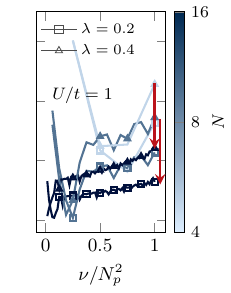}
  }%
  %
  \caption{\label{fig:2rdm:condensate}
    In \subfigref{fig:2rdm:eig-vals} the fragmentation of the~\gls{2RDM} driven by strong correlations is illustrated.
    Increasing the particle number $N$, the formation of plateaus with occupation $n_\nu \sim 1$, composed of $N^2_p= (N/2)^2$ eigenvalues is observed for $U/t > \lambda > 0$.
    The eigenmodes $\hat \varphi^\dagger_\nu$ associated with the plateaus are close to their maximal occupation, which is illustrated in~\subfigref{fig:2rdm:commutators}, where we plot the deviation $\mathcal E$ of the $\hat \varphi^\dagger_\nu$'s commutation relations from hardcore bosonic ones.
    The arrows indicate how the hardcore bosonic character of the $\hat \varphi^\dagger_\nu$'s evolves with increasing $N$ and $U$.
  }
\end{figure}
We now turn to the question of correlation\hyp driven electron pairing and condensation at finite filling fractions.
For that purpose, we diagonalize the~\gls{2RDM}'s matrix elements $\rho_{(ij),(mn)}$ yielding the representation of $\hat\rho^{(2)}$ in terms of its eigenmodes $\hat\varphi^{(\dagger)}_\nu = \sum_{ij} \overbar{v}_{\nu, ij} \hat c^\dagger_{i,\uparrow}\hat c^\dagger_{j,\downarrow}$
\begin{equation}
  \hat\rho^{(2)} = \sum_{\nu} n^\nodagger_\nu \hat\varphi^\dagger_\nu \hat\varphi^\nodagger_\nu \;,
\end{equation}
with $v_{\nu, ij}$ being the $(ij)$th component of the $\nu$th eigenvector of $\rho_{(ij),(mn)}$.
The standard procedure to detect a superfluid or superconducting state, i.e., (quasi) \gls{ODLRO}, in the ground state is to check for the existence of a dominant eigenvalue $n_{\nu_\mathrm{max}}\sim N^\alpha$~\cite{Penrose1956,Yang1962} with $N$ being the number of electrons in the system.
The exponent $\alpha$ is bounded to $\alpha \leq 1/2$ in finite\hyp range interacting, one\hyp dimensional systems due to the Mermin\hyp Wagner theorem~\cite{Mermin1966,Mermin1967}.
Besides the detection of \gls{ODLRO}, recently the possible existence of fragmented condensates has been suggested~\cite{Wietek2022}, a situation in which not only one but several eigenvalues acquire dominant behavior.
In this case, it becomes crucial also to investigate the eigenvectors of $\rho_{(ij),(mn)}$, which can provide insights into the correlation structure of the ground\hyp state wave function.
In ~\cref{fig:2rdm:eig-vals} we show the largest eigenvalues $n_\nu$ of the~\gls{2RDM} in the ground state as a function $\nu$ sorted in descending order, varying the filling $N$ as well as both the electron\hyp electron and electron\hyp phonon interactions, for a system with $L=64$ lattice sites.
For the $U/t=0$ case shown in the left panel, we observe a softening of the dominant eigenvalues when the filling is increased and, crucially, no evidence for an extensively scaling eigenvalue, i.e., no onset of condensation into one mode.
This softening is a precursor for a model instability where we find that for $\lambda = 0.5$ the model becomes unstable due to an unphysical change of the sign of hopping induced by large phonon distortions~\cite{Nocera_2021,supp_mat}.
Upon onset of this unphysical behavior, electrons form localized puddles of charge, i.e. "phase\hyp separate", which translates into a wide distribution of two-particle states in momentum space and thus lead to the loss of the plateau\hyp like structure in the~\gls{2RDM} eigenvalues.
The situation changes upon turning on the repulsive interactions as shown in the middle and right panels for the cases $U/t=0.5$ and $U/t=1$, respectively.
Here, with increasing the filling, we observe a flattening of the largest eigenvalues close to $n_\nu = 1$, which is stabilized up to larger values of $\lambda$ and a sharp drop after the first $(N/2)^2$ eigenvalues.
We interpret this observation as the formation of a maximally fragmented condensate of two\hyp particle bound states which we will justify in the following.
We first note that given the form of the eigenmodes $\hat\varphi^{(\dagger)}_\nu$, there are $(N/2)^2 \equiv N^2_p$ possible, orthogonal two\hyp particle states.
In order to rule out that the observation of exactly $N_p^2$ eigenvalues of magnitude $\sim 1$ is accidental, we calculate the commutation relations of the eigenmodes, which can be written as
\begin{equation}
  \braket{\left[\hat \varphi^\nodagger_\nu, \hat \varphi^\dagger_{\nu^\prime} \right]}
  =
  \delta_{\nu,\nu^\prime} - \sum_{jlm, \sigma} \overbar{v}_{\nu, jm}v_{\nu^\prime, lm} \braket{\hat c^\nodagger_{j,\sigma} \hat c^\dagger_{l,\sigma}} \; .
\end{equation}
For bosonic commutation relations, the second contribution must vanish, which is a necessary condition for the existence of a conventional condensate.
Since the observed occupations already indicate the absence of condensation, we evaluate the relative deviation from the hardcore bosonic case
\begin{equation}
  \mathcal E
  =
  \frac{1}{2 n_\nu}
  \left\lvert
  \delta_{\nu^\noprime,\nu^\prime}(1-2 n_\nu)
  -
  \braket{\left[\hat \varphi^\nodagger_{\nu^\noprime}, \hat \varphi^\dagger_{\nu^\prime} \right]}
  \right\rvert \;,
\end{equation}
shown in~\cref{fig:2rdm:commutators}.
Increasing the repulsive on\hyp site interaction $U/t$, we observe a strong tendency towards hardcore bosonic commutation relations, i.e., $\mathcal E \rightarrow 0$, with increasing particle number, indicated by the arrows in the middle and right panels.
Thus, the observed occupations of the~\gls{2RDM} are close to their maximal value and in particular $n_\nu > 1$ is impossible.
\Cref{fig:2rdm:commutators} suggests that in this regime the~\gls{2RDM} eigenmodes satisfy in the ground state the commutation relations
\begin{equation}
  \braket{\left[\hat \varphi^\nodagger_\nu, \hat \varphi^\dagger_{\nu^\prime} \right]}
  \approx
  \delta_{\nu,\nu^\prime}\left( 1 - \braket{\hat n^{\varphi}_{\nu,\uparrow} } - \braket{ \hat n^{\varphi}_{\nu,\downarrow} } \right)\;,
\end{equation}
where we introduced the density operator $\hat n^{\varphi}_{\nu,\sigma}$ of the eigenmodes.
The $SU(2)$ symmetry of the~\gls{SSH} model suggests the identification $n_\nu = \braket{\hat n^{\varphi}_{\nu,\uparrow}} = \braket{\hat n^{\varphi}_{\nu,\downarrow}}$.
As a consequence, we find that the~\gls{2RDM} can be well\hyp approximated by a mixed state
\begin{equation}
  \hat\rho^{(2)} = \sum_{\nu=1}^{N_p^2} n_\nu \ket{\varphi_\nu}\bra{\varphi_\nu}
\end{equation}
where the two\hyp particle states $\ket{\varphi_\nu}$ describe hardcore bosons.
Note that while the $\ket{\varphi_\nu}$ can also be interpreted as composite bosons~\cite{Rombouts2002,Combescot2008}, here the effective hardcore constraints occur in the space of the~\gls{2RDM}'s eigenmodes labeled by the mode\hyp index $\nu$.
\paragraph*{Cooper\hyp pairing at finite densities}
\begin{figure}
  \centering
  \subfloat[\label{fig:2rdm:correlations:k-space}]{
    \hspace*{-0em}
    \includegraphics{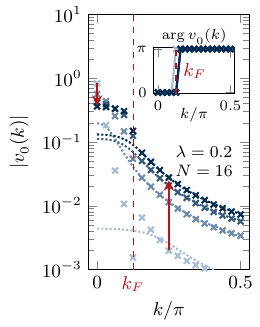}
    \hspace{-1em}
    \includegraphics{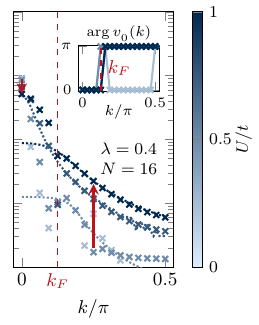}
  }%
  \\
  \subfloat[\label{fig:2rdm:correlations:real-space}]{
    \includegraphics{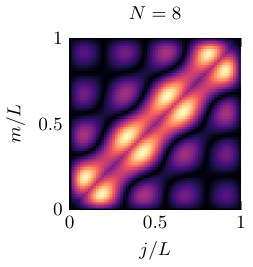}
    \hspace{-1.em}
    \includegraphics{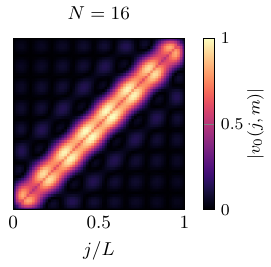}
  }
  \caption{\label{fig:2rdm:correlations}
  Emergence of polaronic Cooper\hyp pairing in the~\gls{2RDM} eigenstate with $\nu=0$ corresponding to the largest eigenvalue.
  In the main plots of~\subfigref{fig:2rdm:correlations:k-space} the momentum\hyp space components $v_\nu(k_\uparrow, k_\downarrow)$ of the~\gls{2RDM}'s eigenmodes are shown along their dominant contributions $k_\uparrow = -k_\downarrow \equiv k$.
  Above the Fermi momentum $k>k_F$ and in the strongly coupled regime $U/t > \lambda > 0$ a Cooper\hyp instability is inferred, where in the insets the phase of the~\gls{BCS}\hyp pairing gap $\Delta_k = v_0(k)$ is shown.
  The dotted lines in the main plots are obtained from fitting $\lvert v_0(k) \rvert$ for momenta with $\Delta_k < 0$, using the~\gls{BCS} wave function.
  Real space components $\lvert v_{\nu,jm}\rvert$ of the eigenmodes are shown in~\subfigref{fig:2rdm:correlations:real-space} in the strongly interacting regime $U/t = 1.0$, $\lambda = 0.4$.
  A characteristic length scale $\pi / k_F=\frac{2L}{N}$ of Cooper\hyp pairing set by $\Delta_{k}<0$ if $k>k_F$ is clearly visible in the form of non\hyp overlapping areas of charge accumulation.
  }
\end{figure}
A physical picture of the maximally fragmented condensate can be established by investigating the correlation structure in the two\hyp particle states $\ket{\varphi_\nu}$.
Taking a Fourier transformation with respect to the spin\hyp up (index $j$) and spin\hyp down (index $m$) components of the~\gls{2RDM}'s eigenstates $v_{\nu, jm}$, yields their momentum\hyp space components $v_\nu(k_\uparrow,k_\downarrow)$.
We find that the different eigenmodes $\ket{\varphi_\nu}$ can be characterized by a finite momentum $k_\nu$ and are dominated by contributions $v_\nu(k_\uparrow,k_\downarrow) \approx v_\nu(k) \delta_{k_\downarrow, -k_\uparrow + k_\nu}$:
\begin{equation}
  \ket{\varphi_\nu} = \frac{1}{L}\sum_{k} v_\nu (k) \hat c^\dagger_{\uparrow,k} \hat c^\dagger_{\downarrow,-k + k_\nu}\ket{\varnothing} \; .
\end{equation}
In general, we observe that the largest eigenvalue $n_0$ has $k_\nu = 0$.
This case is shown in~\cref{fig:2rdm:correlations:k-space} for $N=16$ using $L=64$ lattice sites.
Upon increasing the repulsive interactions, the strong peak at $k=0$ is transformed into a plateau when $U/t>\lambda>0$, decaying algebraically for $k>k_F$.
This is reminiscent of Cooper\hyp pairing and we further corroborate this observation by showing the phase factor of $v_\nu (k)$ in the insets.
In the~\gls{BCS} picture, the amplitudes of the two\hyp particle states are related to the phase factor of the momentum dependent pairing gap
\begin{equation}
  \Delta_k = \lvert \Delta_k \rvert \mathrm e^{\mathrm i \varphi_k} = \lvert \Delta_k \rvert \mathrm e^{\mathrm i \operatorname{arg} v_\nu (k)} \;.
\end{equation}
Here, we can observe a characteristic jump from $\operatorname{arg} v_\nu (k<k_F) = 0$ to $\operatorname{arg} v_\nu (k>k_F) = \pi$.
%
Hence, $\operatorname{arg} v_\nu (k) = \pi$ yields a momentum dependent pairing gap $\Delta_{k>k_F} < 0$, indicating a Cooper instability.
This is supported by fitting the wave functions with the usual \gls{BCS}\hyp type ansatz for $k>k_F$:
\begin{equation}
  v_\nu(k>k_F) = \mathcal N \sqrt{1-\frac{\xi_k}{\sqrt{\xi^2_k + \Delta^2}}}\;,
\end{equation}
with $\xi_k = -2t\cos(k) - \mu$ where $\mu$ is the chemical potential, and $\Delta$ the~\gls{BCS} gap.
We find excellent agreement for the~\gls{BCS}\hyp type ansatz applied to the~\gls{2RDM}'s eigenstates in the strong coupling regime $U/t > \lambda > 0$, shown in~\cref{fig:2rdm:correlations:k-space} as dotted lines.
We emphasize that Cooper\hyp pairing is restricted to a length scale set by $a \sim 1/k_F$.
In real space, this translates into spatially non-overlapping, correlated areas of charge accumulations, each of which can be understood as region with dominant pairing correlations, as shown in~\cref{fig:2rdm:correlations:real-space} for the case $L=64$, $U/t=1$ and $\lambda = 0.4$ at fillings $N=8$ and $N=16$.
Notably, for $\nu>0$ we always find finite\hyp momentum Cooper pairing, which follows from the fact that the different eigenstates of the~\gls{2RDM} have to be orthogonal but can not be populated with more than one pair of electrons (i.e., effective hardcore constraint).
\paragraph*{Discussion and Outlook\label{sec:conclusion}}
Our results have important consequences regarding the conventional picture of bipolaronic superconductivity in the adiabatic limit.
We demonstrated that for the one\hyp dimensional~\gls{SSH}\hyp model, molecular bipolarons are unstable in the physically relevant regime.
Our analysis of the~\gls{2RDM} indicates that at strong interactions $U/t > \lambda > 0$, polarons form pairs with a large spatial coherence length, reminiscent of Cooper pairing.
Therefore, the ground state at finite filling fractions is not a superfluid of molecular bipolarons, but rather a strongly correlated mixture of bipolaron\hyp like polaronic pairs with finite momenta in a state, characterized as a maximally fragmented condensate.
We also point out that at finite fillings, the interplay between strong repulsive interactions and comparably weak electron\hyp phonon couplings is what yields the most pronounced pairing, a setting that is more typical of actual physical systems.
A possible physical mechanism behind these findings may be a tendency of the off\hyp diagonal electron\hyp phonon coupling to generate pair\hyp density wave correlations.
In fact, there is recent evidence that such density fluctuations could generate fragmentation~\cite{Baldelli2023}, an avenue we plan to explore in future work.
These results call for further unbiased investigations of the phase diagram of electron\hyp phonon systems with off\hyp diagonal coupling in order to establish possible pairing mechanisms in physically realistic systems.
\section*{Acknowledgments\label{sec:acknowledgments}}
We thank Alexander Wietek for inspiring discussions and Lode Pollet, Mona Berciu and Andrew Millis for comments on the manuscript.
M. G., T. B., U. S. and S. P. acknowledge support by the Deutsche Forschungsgemeinschaft (DFG, German Research Foundation) under Germany’s Excellence Strategy-426 EXC-2111-390814868.
J. S. acknowledges support from the Gordon and Betty Moore Foundation’s EPiQS Initiative through Grant GBMF8686 at Stanford University.

\appendix

\section{Singlon\hyp Doublon decomposition\label{sec:supp:decomposition}}
In order to analyze the quasi\hyp particles of the~\gls{SSH} model~\cite{SSH_1979}, it is convenient to represent the Hamiltonian in terms of momentum space operators.
For that purpose, we introduce the Fourier transformed electronic spin\hyp $\sigma$ $\hat c^{(\dagger)}_{k,\sigma}$ and $\hat a^{(\dagger)}_k$ annihilation (creation) operators
\begin{align}
    \hat c^\nodagger_{k^\noprime_n,\sigma} &= \frac{1}{\sqrt L} \sum_j \mathrm e^{-\mathrm i k^\noprime_n r_j} \hat c^\nodagger_{j,\sigma} \; , \\
    \hat a^\nodagger_{k^\noprime_n} &= \frac{1}{\sqrt L} \sum_j \mathrm e^{-\mathrm i k^\noprime_n r_j} \hat a^\nodagger_{j} \; ,
\end{align}
where $r_j = j\cdot a$ denotes the position of the $j$th lattice site and $a$ is the lattice constant.
Here, the discrete momenta $k_n$ are elements of the Brillouin zone $k_n = \frac{\pi}{L} n$ with $n=-L, -L+1, \ldots, L-1$.
For notational convenience we drop the discrete index $n$: $k_n \rightarrow k$ and implicitely perform summations over momentum space indices $k,q$ with respect to their allowed ranges depending on the number of lattice sites $L$.
The momentum space representation of the phononic displacement operator $\sqrt{2 M \Omega} \hat X^\nodagger_j = \hat a^\dagger_j + \hat a^\nodagger_j$ with the oscillator mass $M$ can then be expressed as
\begin{equation}
    \hat X^\nodagger_k = \frac{1}{L} \sum_j \mathrm{e}^{-\mathrm i kr_j} \hat X^\nodagger_j \;,
\end{equation}
which satisfies $\hat X^\dagger_k = \hat X^\nodagger_{-k}$, and we set $M\equiv 1$.
Considering $U=0$ for brevity, the~\gls{SSH}\hyp model Hamiltonian in momentum space is given by
\begin{align}
    \hat H
    &=
    -\sum_{k,\sigma} \varepsilon_k \hat c^\dagger_{k,\sigma} \hat c^\nodagger_{k,\sigma}
    +
    \Omega \sum_k \hat a^\dagger_k \hat a^\nodagger_k \notag \\
    &\phantom{=}\quad
    -\sum_{k,\sigma}\sum_{q} \left( V_{k,k-q} \hat c^\dagger_{k,\sigma} \hat c^\nodagger_{k-q,\sigma} \hat X^\nodagger_q + \mathrm{h.c.} \right) \; ,
\end{align}
where we introduced $\varepsilon_k = 2t\cos(k) - \mu$ and the Fourier\hyp transformed electron\hyp phonon coupling
\begin{equation}
    V_{k,q} = \mathrm{i}\alpha\sqrt{\frac{2\Omega}{L}} \left(\sin(k) - \sin(q) \right) \;. \label{eq:supp:el-ph-coupling:k-space}
\end{equation}
Note that for brevity the chemical potential $\mu$ has been supressed in the main text.
We now introduce strictly single\hyp and two\hyp particle electronic operators
\begin{align}
    \hat s^{(\dagger)}_{j,\sigma} &= \left[\hat c^\nodagger_{j,\sigma} (1-\hat n^\dagger_{j,\overbar \sigma}) \right]^{(\dagger)} \; , \\
    \hat d^{(\dagger)}_j &= \left[\hat c^\nodagger_{j,\downarrow} \hat c^\nodagger_{j,\uparrow} \right]^{(\dagger)} \; ,
\end{align}
where we defined the spin\hyp $\sigma$ density operators $\hat n^\nodagger_{j,\sigma} = \hat c^\dagger_{j,\sigma} \hat c^\nodagger_{j,\sigma}$ and denoted spin conjugation by an overbar
\begin{equation}
    \overbar \sigma = \begin{cases} \downarrow\;, &\text{if $\sigma=\uparrow$,} \\ \uparrow\;, &\text{if $\sigma=\downarrow$.} \end{cases}
\end{equation}
These operators obey (anti\hyp) commutation relations
\begin{equation}
    \left\{ \hat s^\nodagger_{j,\sigma}, \hat s^\nodagger_{j^\prime,\sigma^\prime} \right\} = 0 \;,
    \quad \text{and} \quad
    \left[ \hat d^\nodagger_j, \hat d^\nodagger_{j^\prime} \right] = 0 \; ,
\end{equation}
as well as
\begin{align}
    \left\{ \hat s^\nodagger_{j,\sigma}, \hat s^\dagger_{j^\prime,\sigma^\prime} \right\}
    &=
    \delta_{j,j^\prime} \delta_{\sigma, \sigma^\prime} (1-\hat n^\nodagger_{j,\overbar \sigma})
    \\
    \left[ \hat d^\nodagger_j, \hat d^\dagger_{j^\prime} \right]
    &=
    \delta_{j,j^\prime} \mathrm{e}^{\mathrm{i} \pi \hat n^d_j} \;,
\end{align}
with $\hat n^d_j = \hat d^\dagger_j \hat d^\nodagger_j$.
Their Fourier representation is defined via
\begin{align}
    \hat s^\nodagger_{k,\sigma} &= \frac{1}{\sqrt L} \sum_j \mathrm{e}^{-\mathrm{i} k r_j} \hat s^\nodagger_{j,\sigma} \;, \\
    \hat d^\nodagger_{k} &= \frac{1}{\sqrt L} \sum_j \mathrm{e}^{-\mathrm{i} k r_j} \hat d^\nodagger_{j} \;.
\end{align}
Finally, the inverse representation in terms of electronic operators is given by
\begin{align}
    \hat c^\nodagger_{j,\sigma} &= \hat s^\nodagger_{j,\sigma} + \sgn(\sigma) \hat s^\dagger_{j,\overbar \sigma} \hat d^\nodagger_j \; , \\
    \hat c^\dagger_{j,\sigma} &= \hat s^\dagger_{j,\sigma} + \sgn(\sigma) \hat d^\dagger_j \hat s^\nodagger_{j,\overbar \sigma} \; ,
\end{align}
and their Fourier expansion is expressed as
\begin{align}
    \hat c^\nodagger_{k,\sigma} &= \hat s^\nodagger_{k,\sigma} + \frac{\sgn(\sigma)}{\sqrt L} \sum_q \hat s^\dagger_{q-k,\overbar \sigma} \hat d^\nodagger_q \; , \\
    \hat c^\dagger_{k,\sigma} &= \hat s^\dagger_{k,\sigma} + \frac{\sgn(\sigma)}{\sqrt L} \sum_q \hat d^\dagger_q \hat s^\nodagger_{q-k,\overbar \sigma} \; .
\end{align}
Using these definitions as well as $\hat D^\nodagger_{k,\sigma} = \frac{1}{\sqrt L}\sum\limits_q \hat s^\dagger_{q-k,\overbar \sigma} \hat d^\nodagger_q$, the~\gls{SSH}\hyp model can be decomposed into strictly single\hyp and two\hyp particle terms and contributions $\propto \hat D^{(\dagger)}_{k,\sigma}$ that describe scattering processes between them.
We obtain the decompositions
\onecolumngrid

\begin{align}
    \sum_{k,\sigma} \varepsilon_k \hat c^\dagger_{k,\sigma} \hat c^\nodagger_{k,\sigma}
    &=
    \sum_{k,\sigma} \varepsilon_k
    \left\{
        \hat s^\dagger_{k,\sigma} \hat s^\nodagger_{k,\sigma}
        +
        \hat D^\dagger_{k,\sigma} \hat D^\nodagger_{k,\sigma}
        +
        \sgn(\sigma) \left(\hat s^\dagger_{k,\sigma} \hat D^\nodagger_{k,\sigma}  + \mathrm{h.c.} \right)   
    \right\} \\
    \sum_\sigma\sum_{k,q} V_{k,k-q} \hat c^\dagger_{k,\sigma} \hat c^\nodagger_{k-q,\sigma} \hat X^\nodagger_q
    &=
    \sum_{k,\sigma}\sum_{q} V_{k,k-q}
    \left\{
        \hat s^\dagger_{k,\sigma} \hat s^\nodagger_{k-q,\sigma}
        +
        \hat D^\dagger_{k,\sigma} \hat D^\nodagger_{k-q,\sigma}
        +
        \sgn(\sigma) \left( \hat s^\dagger_{k,\sigma} \hat D^\nodagger_{k-q,\sigma} + \mathrm{h.c.} \right)
    \right\} \hat X^\nodagger_q \; .
\end{align}
\twocolumngrid
Let us briefly comment on the different operators and terms.
The operators $\hat s^{(\dagger)}_{k,\sigma}$ describe spin\hyp $\sigma$ fermions (singlons) that are composed of a superposition of spin\hyp $\sigma$ electrons, in which the electron solely occupies the lattice site, i.e., there are no double occupancies.
Conversely, the operators $\hat d^{(\dagger)}_k$ describe a superposition of hardcore bosons composed of electronic double occupancies (doublons).
Let us now assume that there are stable on\hyp site bipolarons, glued together by the electron\hyp phonon interaction.
Apparently, these bipolarons are then described by doublons dressed with phonons and if there are electrons that are not bound into on\hyp site bipolarons, they must be described by singlons.
The bilinear contributions $\sum_\sigma \hat D^\dagger_{k,\sigma} \hat D^\nodagger_{q,\sigma} \equiv \hat D^\nodagger_k \hat D^\nodagger_q$ describe scattering processes between doublons and singlons where the total numbers of both species
\begin{equation}
    \hat N_{s,\sigma} = \sum_j \hat s^\dagger_{j,\sigma} \hat s^\nodagger_{j,\sigma} \;,
    \quad
    \hat N_{d} = \sum_j \hat d^\dagger_{j} \hat d^\nodagger_{j}
\end{equation}
are conserved.
Finally, there are processes $\hat s^\dagger_{k,\sigma} \hat D^\nodagger_{q,\sigma}$ and $\hat D^\dagger_{k,\sigma} \hat s^\nodagger_{q,\sigma}$ converting doublons into singlons and vice versa, assisted by the annihilation/creation of an additional singlon with opposite spin $\overbar \sigma$.
\section{Instability of on\hyp site bipolarons\label{sec:supp:bipolaron-instability}}
From the decomposition in~\cref{sec:supp:decomposition} we can analyze the stability of an on\hyp site bipolaron, described by dressed doublons.
In the $N=2$ particle sector with total spin $S^z=0$, if there is such a stable quasi particle, the eigenstates have the form
\begin{equation}
    \ket{\psi^d_k} = \sum_q \psi^d_{k}(q) \hat d^\dagger_q \ket{\varnothing}_\mathrm{el}\ket{a^\nodagger_k(q)}_\mathrm{ph} \label{eq:eigenstates:on-site-bipolarons}
\end{equation}
where $\ket{\varnothing}$ is the electronic vacuum, $\ket{a_q(k)}$ a phonon configuration characterized by the momenta $k$ and $q$ and $\psi^d_k(q) \in \mathbb C$.
For this ansatz to be an eigenstate, all matrix elements describing pure singlons and singlon\hyp doublon conversion must vanish such that it suffices to consider only the matrix elements of the Hamiltonian containing bilinear contributions $\hat D^\nodagger_k \hat D^\nodagger_q$:
\begin{equation}
    \hat H_{D-D}
    =
    \sum_{k} \left(\varepsilon_k \hat D^\dagger_k \hat D^\nodagger_k + \sum_q \left(V_{k,k-q} \hat D^\dagger_k \hat D^\nodagger_{k-q} \hat X^\nodagger_q + \mathrm{h.c.} \right) \right) \; .
\end{equation}
Calculating the matrix elements w.r.t. the electronic states we find
\onecolumngrid

\begin{align}
    {\vphantom{\ket{\varnothing}}}_\mathrm{el}\braket{\varnothing|\hat d^\nodagger_k  \hat H_{D-D} \hat d^\dagger_{k^\prime} | \varnothing}_\mathrm{el}
    &=
    \frac{2}{L} \left\{ \delta_{k,k^\prime} \sum_q \varepsilon_q + \sum_q\left(V_{q,q-(k-k^\prime)} \hat X^\nodagger_{k-k^\prime} + \mathrm{h.c.} \right) \right\} + \Omega\sum_q \hat a^\dagger_q \hat a^\nodagger_q + U \notag \\
    &\stackrel{L\rightarrow\infty}{=}
    2\mu \delta_{k,k^\prime} + \frac{1}{\pi}\left(\hat X^\nodagger_{k-k^\prime} - \hat X^\dagger_{k-k^\prime} \right) \int_0^{2\pi} V_{q,q-(k-k^\prime)} \mathrm d q  + \Omega \hat N^\nodagger_a + U \; ,
\end{align}
\twocolumngrid
with the total phonon number $\hat N_a$.
The $q$\hyp integral over the electron\hyp phonon interaction vanishes, which follows immediately from~\cref{eq:supp:el-ph-coupling:k-space} where we can integrate both summands independently.
Since $\braket{\hat N^\nodagger_a} \geq 0$ it follows that eigenvalues $E^d_k$ belonging to eigenstates of the form~\cref{eq:eigenstates:on-site-bipolarons}, if existant at all, satisfy $E^d_k \geq \mu + U$.
Hence, there are no bound on\hyp site bipolaron states in the two\hyp particle sector, i.e., bipolaronic bound states must be delocalized over at least two lattice sites.
In the bipolaronic spectral function we indeed observe these states as incoherent signal at energies $\omega\sim \Omega+U$ which is consistent with the choice $\mu=0$ for our calculations.
However, this band exhibits a small dispersion that can be addressed to scatterings into polarons, which we neglected in the above calculation.
\section{Projected purified matrix-product states\label{sec:supp:pp-mps}}
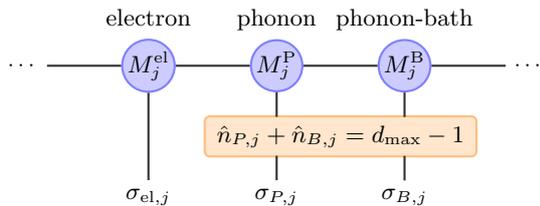
\begin{figure}
  \centering
  \tikzsetnextfilename{ssh_mps}
  \begin{tikzpicture}
    \begin{scope}[node distance = 1.0em and 3.0em]
    \node [site] (site1) [label=above:$\vphantom{p}$electron] {\small $M^\mathrm{el}_{j}$};
    \node [site, right=of site1] (site2) [label=above:phonon] {\small $M^\mathrm{P}_{j}$};
    \node [site, right=of site2] (site3) [label=above:phonon-bath] {\small $M^\mathrm{B}_{j}$};
    \node [nosite, below=of site2] (op2) {};
    \node [nosite, below=of site3] (op3) {};
    \node [gate, fit=(op2)(op3),minimum width=11em,label={center:$\hat n_{P,j}+\hat n_{B,j}=d_\mathrm{max}-1$}] {};
    \node [ghost] [below=of op2] (below2) {\small $\sigma_{P,j}$};
    \node [ghost] [below=of op3] (below3) {\small $\sigma_{B,j}$};
    \node [ghost] at (below2-|site1) (below1) {\small $\sigma_{\mathrm{el},j}$};
    \node [ghost] [left=of site1] (left1) {$\cdots$};
    \node [ghost] [right=of site3] (right3) {$\cdots$};

    \draw [thick, draw=black!80] (site1) to (below1);
    \draw [thick, draw=black!80] (site2) to (op2);
    \draw [thick, draw=black!80] (site3) to (op3);
    \draw [thick, draw=black!80] (op2) to (below2);
    \draw [thick, draw=black!80] (op3) to (below3);
    \draw [thick, draw=black!80] (site1) to (site2);
    \draw [thick, draw=black!80] (site2) to (site3);
    \draw [thick, draw=black!80] (site1) to (left1);
    \draw [thick, draw=black!80] (site3) to (right3);
    \end{scope}
  \end{tikzpicture}
  \caption{
    \label{fig:ssh:mps}
    Decomposition of the unit cell, containing an electronic degree of freedom $\ket{\sigma_{\mathrm{el},j}}$, a phononic degree of freedom $\ket{\sigma_{P,j}}$ and the corresponding phononic bath site $\ket{\sigma_{B,j}}$.
    The \acrshort{PP} gauge constraint is indicated as two\hyp site gate, which, in practice, is absorbed into the construction of the Hamiltonian~\cite{Stolpp2021}
  }
\end{figure}
One of the main advantageous of~\gls{TNS}\hyp based techniques is their capability to provide a controlled approximation of the total many\hyp body wavefunction.
For the case of~\gls{MPSs}, this representation is obtained from an ansatz class for the wavefunction's coeficients of the form
\begin{equation}
  c_{\sigma_1,\ldots,\sigma_L} = \sum_{\sigma_1,\ldots,\sigma_L} M^{\sigma_1} \cdots M^{\sigma_L} \ket{\sigma_1,\ldots,\sigma_L} \; ,
\end{equation}
where the $\sigma_j = 0, \ldots,d_j-1$ denote the $L\in \mathbb N$ local degrees of freedom, spanning the overall many\hyp body Hilbert space, and $M^{\sigma_j} \in \mathbb C^{m_{j-1}\times m_j}$ are complex matrices with bond dimensions $m_j$, encoding the wavefunction.
For systems with small local Hilbert space dimensions $d_j$, there are efficient contraction and compression schemes but the situation changes drastically, if some of the local degrees of freedoms describe harmonic oscillators where in principle $d_j = \infty$.
In that cases one typically introduces a finite cutoff dimension $d_\mathrm{max}$, which, however must be chosen carefully such that the physical properties are not affected by the artifical truncation.
Clearly, selecting a proper value $d_\mathrm{max}$ depends on the particular basis chosen to represent the bosonic modes and different ways have been explored in the past~\cite{Jeckelmann1998,Brockt2015,Koehler2021,Stolpp2021}.
In this work, we use~\gls{PP-MPSs} which allow for a dynamical adoption of the number of actually used local basis states $d_\mathrm{eff} \leq d_\mathrm{max}$, given a suitably large value of $d_\mathrm{max}$.
The key idea is to introduce for each bosonic lattice site $\ket{\sigma_{P,j}}$ an artifical bath site $\ket{\sigma_{B,j}}$ that acts as reservoir for the physical bosonic excitations, such that for each pair of physical and bath site, the overall number of bosonic excitations is conserved: $\hat n_{P,j} + \hat n_{B,j} = d_\mathrm{max}-1$.
This gauge constraint introduces a global $U(1)$\hyp symmetry for the bosonic degrees of freedom, allowing for a drastic reduction of the numerical complexity, in particular if parallelization is exploited~\cite{Stolpp2021,Mardazad2021}.
Furthermore, it can be shown that for a bipartition of the overall system between bosonic physical and bath sites, the Schmidt values can be identified with the bosonic excitation probabilities, allowing for a controlled truncation of irrelevant bosonic basis states.

\section{Time Evolution Methods\label{sec:supp:methods}}
\begin{figure}[h]
  \centering
  \hspace{-10mm}
  \includegraphics[width=\linewidth]{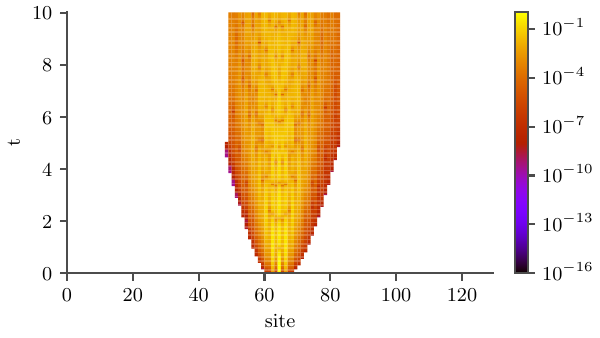}
  \caption{\label{fig:chu_2tdvp}
    Light cone for an electronic excitation at $\Omega=0.4$ and $\lambda=0.5$ using~\gls{GSE} and~\gls{2TDVP}.}
\end{figure}
\begin{figure}[h]
  \centering
  \hspace{-10mm}
  \includegraphics[width=\linewidth]{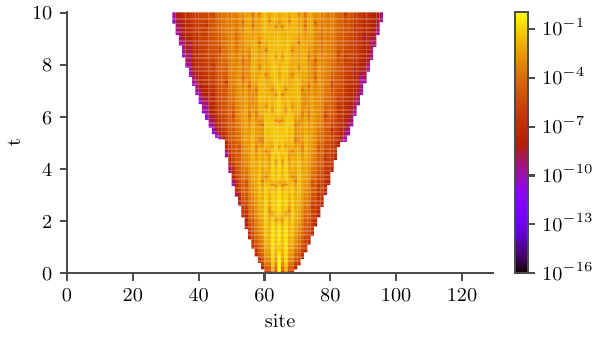}
  \caption{\label{fig:chu_lse}
    Light cone for an electronic excitation at $\Omega=0.4$ and $\lambda=0.5$ using~\gls{GSE} and~\gls{LSE}.}
\end{figure}
Due to their unfavourable scaling in the local dimension of our Hilbert space, the commonly used~\gls{TNS} time evolution methods \cite{Paeckel2019} ~\gls{TDVP}, in its two-site variant, \cite{Haegeman_2011,Haegeman_2016} or ~\gls{TEBD} \cite{Daley_2004,Vidal2004,White2004} perform notoriously poor for the simulation of electron-phonon systems.
Recent developments brought forth subspace expansion extensions for~\gls{1TDVP} that scale with $d^2$ instead of $d^3$, whilst still conserving the ability to dynamically grow the bond dimension\cite{Yang_2020prb,Yang_2020}.
The ~\gls{GSE} builds a small Krylov subspace of typically $3-5$ Krylov vectors that can be used to enrich the state.
As the construction of these Krylov Vectors is rather expensive, we result to the ~\gls{LSE} for most of our calculation. 
LSE aims to enrich the state with a local expansion tensor. We are using the expansion scheme introduced by Hubig et al. \cite{Hubig_2015}: 
\begin{equation}
    E_{i} = L_{i-1}M_{i}W_{i}
\end{equation}
where $L_{i-1}$ are the left contractions, $M_{i}$ is the active site tensor and $W_{i}$ the corresponding MPO. 
The contractions can be build iteratively during a TDVP sweep, $L_{i}=L_{i-1}A^{\dagger}_{i}W^{\nodagger}_{i}A^{\nodagger}_{i}$ with $L_{0}=1$.
It not only scales better than~\gls{2TDVP} in the local dimension it is also more stable in regards to small bond dimensions \cite{Yang_2020,Paeckel2019}.
This is crucial for our calculations as we are calculating the dynamics of excitations on a product state.
\Cref{fig:chu_2tdvp} shows the light cone of an electronic excitation at $\Omega=0.4$ simulated using~\gls{2TDVP}.
As we can see, the expansion of the light cone freezes and the excitation is trapped in a system of finite size. 
LSE, however, manages to accurately continue spread theof the excitation.
\subsection*{Local Subspace Expansion - LSE}
In the following we will give a step wise description of a left-to-right MPS sweep of the LSE.
The update scheme of the $i$th site tensor is\cite{Yang_2020}: 
\begin{enumerate}
    \item time evolve the site tensor $M_i(t)\rightarrow M_i(t+\frac{\delta t}{2})$
    \item truncate the site tensor $M_i$ using a SVD and obtain the decomposition $M_{i}=B_{i}S_{i}A_{i}$.
    \item evaluate heuristically whether to expand, renormalize $S_{i}$ and build bond tensor $D_i=S_iA_i$
    \begin{enumerate}
        \item  if no expansion is needed update site tensor $M_i=B_i$; jump to step \ref{backevo}
        \item else, build the projector onto the orthogonal compliment of $M_{i}$, $P_i=\mathbb{1}-B^{\nodagger}_i B_i^{\dagger}$
        \end{enumerate}
    \item build the expansion tensor $ E_{i} = \alpha_{i}L_{i-1}M_{i}W_{i}$
    \item project it onto the orthogonal compliment $E^{\perp}_i=E_{i}P_{i}$
    \begin{enumerate}
        \item  if the norm is of $E^{\perp}_i$ is zero, update site tensor $M_i=B_i$; jump to step \ref{backevo}
        \item else, truncate the expansion tensor and obtain the decomposition $E^{\perp}_{i}=B^{\perp}_{i}S^{\perp}_{i}A^{\perp}_{i}$
        \end{enumerate}
    \item enlarge $B_{i}$ via direct sum as $\Tilde{B}_i=\mqty[B^{\nodagger}_{i} & B^{\perp}_{i}]$
    \item build new expanded bond tensor via $D_{i}=\Tilde{B}^{\dagger}_i M^{\nodagger}_i$
    \item calculate new left contractions $L_{i}=L_{i-1}A^{\dagger}_{i}W^{\nodagger}_{i}A^{\nodagger}_{i}$
    \item back evolve bond tensor $D_{i}(t+\frac{\delta t}{2}) \rightarrow D_i(t)$ \label{backevo}
    \item multiply $D_i$ into the next site tensor $B_{i+1}$
\end{enumerate}
A further advantage of this method over~\gls{2TDVP} is, that it is not plagued by stability issues at small bond dimensions\cite{Paeckel2019}.
This turns out to be crucial for our simulations as we are calculating the dynamics of an excitation on a product state.
As shown in \cref{fig:chu_2tdvp},~\gls{2TDVP} fails to accurately describe the spread of the excitation after we switch the time evolution method from~\gls{GSE} to~\gls{2TDVP} at $T=5$.
\subsection*{Extracting (Bi\hyp)Polaronic Dispersion Relation from Spectral Functions}
\begin{figure}[ht!]
    \subfloat[\label{fig:spec:low-energy:polaron}]{
        \centering
        \includegraphics{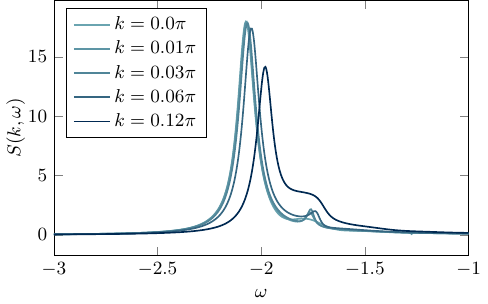}
    }\\
    \subfloat[\label{fig:spec:low-energy:bipolaron}]{
        \centering
        \includegraphics{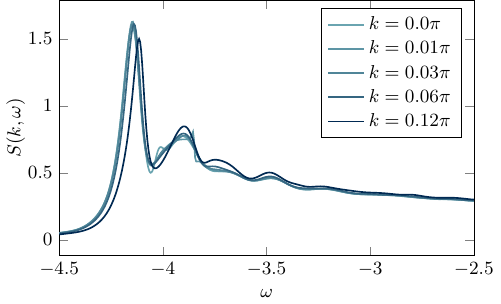}
    }\\
    \subfloat[\label{fig:binding-energies:bipolarons:w-1}]{
        \centering
        \includegraphics{figures/low_energy_spec_bipolaron.pdf}
    }
    \caption{\label{fig:spec:low-energy}
    Momentum cuts of spectral functions calculated for $U/t=0$ and $\lambda=\nicefrac{1}{2}$.
    The polaron spectral functions are shown in~\subfigref{fig:spec:low-energy:polaron}, the bipolaronic spectral functions in~\subfigref{fig:spec:low-energy:bipolaron}.
    Dots are indicating the position of the maximum of the energetically lowest peak, from which the polaronic/bipolaronic dispersion relation is reconstructed.
    In~\subfigref{fig:binding-energies:bipolarons:w-1}, the bipolaronic binding energy at larger phonon frequencies $\Omega/t = 1$ is shown.
    Previous investigations already reported the formation of bipolarons in the anti- adiabatic regime, which is confirmed by our observations of large bipolaronic binding energies $\lvert\Delta_\mathrm{BP}(k = 0)/t\rvert$.
    }
\end{figure}
We calculated the time\hyp dependent, (bi\hyp)polaron Greens functions
\begin{align}
  G_\mathrm{P}(j,t) &= \braket{\varnothing | \hat c^\nodagger_{j,\sigma}(t) \hat c^\dagger_{\nicefrac{L}{2},\sigma}(0) | \varnothing}\quad\text{polaron,} \label{eq:gf:polaron}\\
  G_\mathrm{BP}(j,t) &= \braket{\varnothing | \hat d^\nodagger_j(t) \hat d^\dagger_{\nicefrac{L}{2}}(0) | \varnothing}\quad\text{bipolaron,} \label{eq:gf:bipolaron}
\end{align}
where $\ket{\varnothing}$ denotes the combined electron\hyp phonon vacuum state, $\hat c^{(\dagger)}_{j,\sigma}$ are spin\hyp$\sigma$ electronic annihilation (creation) operators and $\hat d^{(\dagger)}_j = \hat c^\dagger_{j,\uparrow}\hat c^\nodagger_{j,\downarrow}$.
Taking two Fourier transformations then yields the momentum resolved spectral functions (with $\alpha=\mathrm{B,BP}$)
\begin{equation}
  S_\alpha(k,\omega) = \frac{\pi}{T\sqrt L} \sum_{j,n} \mathrm{e}^{-\mathrm{i}(k\cdot j - \omega \cdot t_n) - \eta t_n} G_{\alpha}(j,t_n) \;, \label{eq:sf}
\end{equation}
where the $j$\hyp sum is over all lattice sites.
In~\cref{fig:spec:low-energy} we show momentum cuts of the obtained (bi\hyp)polaronic spectral functions, exemplarically.
The dispersion relation is then extracted by determining the position of the low\hyp energy peaks (indicated as dots) $\varepsilon_\mathrm{P/BP}(k)$.
We fit their positions around their minimum value $\varepsilon^0_\mathrm{P/BP}$ using a quadratic ansatz
\begin{equation}
    \varepsilon_\mathrm{P/BP}(k)=\varepsilon_\mathrm{P/BP}^0+\frac{k^2}{2m^\noprime_\mathrm{eff}}\;,
\end{equation}
with the quasi\hyp particle effective mass $m^\noprime_\mathrm{eff} = \left. \frac{\partial^2 E}{\partial k^2} \right\vert_{k=0}$
For the considered couplings in the adiabatic limit, we always find $\varepsilon^0_\mathrm{P/BP}$ to be located at $k=0$ for both, polarons ($\mathrm P$) and bipolarons ($\mathrm{BP}$).
Previous works already studied the properties of bipolaronic quasi\hyp particles in the anti\hyp adiabatic regime.
In order to connect to these investigations, in~\cref{fig:binding-energies:bipolarons:w-1} we show the bipolaronic binding energies $\Delta_\mathrm{BP}(k = 0) = \varepsilon_\mathrm{BP}(0) - 2\varepsilon_\mathrm{P}(0)$ at larger phonon frequencies $\Omega/t = 1$.
The observation of tightly bound bipolarons with binding energies whose absolute value increases with the effective electron\hyp phonon coupling is in perfect agreement with previous findings~\cite{Sous2018}.
\section{Ground\hyp state results in the low\hyp density limit}
In order to supplement our findings we performed ground\hyp state calculations for the~\gls{SSH}\hyp model in the dilute limit $n=N/2L<0.1$.
In the adiabatic limit, these calculations are challenging due to low energy costs to create phononic excitations.
For that purpose we chose a large number of maximally allowed local phononic excitations $n_\mathrm{ph,max}=63$ which we found to ensure convergence also for the largest systems studied, with up to $L=256$ lattice sites.
For the maximally allowed truncated weight we chose $\delta = 10^{-10}$, for which a maximum bond dimension of $m=2048$ states was sufficient.
Using both, single\hyp and two\hyp site~\gls{MPS} solvers~\cite{Hubig_2015} we converged the ground\hyp state calculations to a relative precision of $\Delta E_0/E_0 = 10^{-8}$ where $\Delta E_0$ denotes the change in the ground\hyp state energy $E_0$ in the last sweep.
However, these parameters need to be taken with a grain of salt, since the~\gls{SSH}\hyp model can become unstable in certain parameter regimes, as discussed below.
Once that unphysical regime is entered, phase separation sets in, generating a large number of excited states that are energetically located very close above the ground state.
In that situation, and using open boundary conditions, the simulations are prone to get stuck in such local minima, which we diagnosed by investigating either the symmetry of local observables such as the electron density, or the (staggered) phonon displacements.
\subsection*{Stability of the \acrshort{SSH}\hyp Model at Strong Electron\hyp Phonon Couplings}
\begin{figure}[ht!]
    \vspace*{1.5em}
    \subfloat[\label{fig:ssh-stability:U-0p0}]{
        \centering
        \def\currentUcol{1}
        \includegraphics{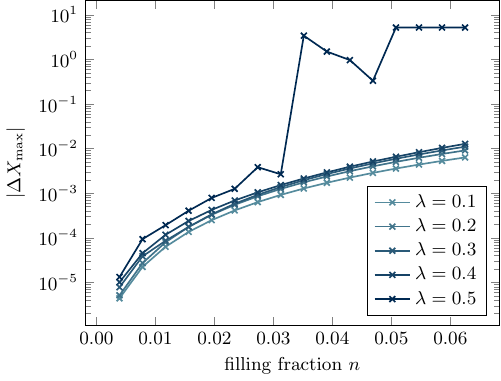}
    }\\
    \subfloat[\label{fig:ssh-stability:U-0p25}]{
        \centering
        \def\currentUcol{2}
        \includegraphics{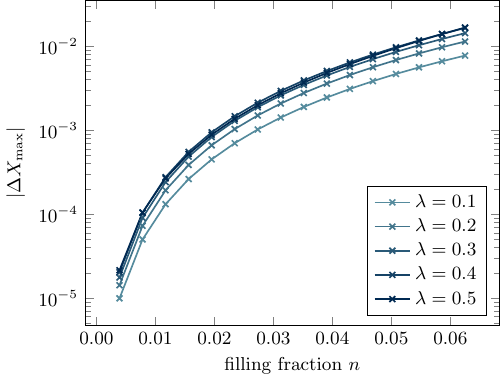}
    }
    \caption{\label{fig:ssh-stability}
    Minimally observed staggered magnetization $\Delta X_\mathrm{max} =  \max_{j\in[0,L-1]} \left(\braket{\hat X_j}-\braket{\hat X_{j+1}}\right)$ as a function of the filling fraction $n=N/L$ and the electron\hyp phonon coupling $\lambda$.
    Values $\lvert \Delta X_\mathrm{max} \rvert>1$ are indicating lattice instabilities and thus a breakdown of the physicality of the~\gls{SSH} model.
    We show exemplary results for $U/t=0$ in~\subfigref{fig:ssh-stability:U-0p0} and $U/t=\nicefrac{1}{4}$ in~\subfigref{fig:ssh-stability:U-0p25} for a system size of $L=256$.
    }
\end{figure}
In the~\gls{SSH} model, the electron\hyp phonon coupling is assumed to be strongly supressed at equilibrium lattice distances $\lvert i-j \rvert > 1$ and therefore neglected.
This approximation can cause unphysical behavior that is reflected in a sign\hyp change of the electron hopping and signalled by displacements $-(\braket{\hat X_j}-\braket{\hat X_{j+1}}) > 1$, i.e., displacements are so large that the atomic coordinates are moved across each other~\cite{Barford_2006}.
Note that in previous studies, due to numerical limitations, even in the non\hyp adiabatic regime, this unphysical behavior was regularized by introducing a rather sharp cutoff in the maximally allowed number of phononic excitations per site~\cite{Barford_2006}.
While this can be justified at large phonon frequencies $\Omega\sim t$, in the adiabatic limit we observe very large numbers of $\sim 30$ excitations per lattice site, which are automatically chosen by the truncation scheme of our~\gls{PP-DMRG} solver, and are numerically controlled w.r.t. the truncated weight~\cite{Koehler2021,Stolpp2021}.
In order to check our results for physical consistency, we performed ground\hyp state calculations and monitored the maximally occuring staggered displacements
\begin{equation}
    \Delta X_\mathrm{max} =  \max_{j\in[0,L-1]} \left(\braket{\hat X_j}-\braket{\hat X_{j+1}}\right) \;.
\end{equation}
Results are shown exemplary for the cases $U/t=0$ in~\cref{fig:ssh-stability:U-0p0} and $U/t=\nicefrac{1}{4}$ in~\cref{fig:ssh-stability:U-0p25}.
For the considered electron\hyp phonon coupling ranges $\lambda \in [0,\nicefrac{1}{2}]$, we observed instabilities only in case of vanishing electron\hyp electron interactions and for filling fractions $n>0.025$, at the largest coupling strength $\lambda=\nicefrac{1}{2}$.

\bibliography{literature}
\end{document}